\documentclass[%
 reprint,
superscriptaddress,
nofootinbib,
 amsmath,amssymb,
 aps,
]{revtex4-2}

\usepackage{graphicx}
\usepackage{dcolumn}
\usepackage{bm}
\usepackage{svg}
\usepackage[frozencache,cachedir=.]{minted}
\usepackage{amsthm}
\newtheorem{example}{Example}

\AtBeginDocument{}

\newcommand{\ket}[1] {\ensuremath{| #1 \rangle}}
\newcommand{\tquote}[1]{``#1''}

\begin{document}

\preprint{APS/123-QED}

\title{MQT Qudits: \\
A Software Framework for Mixed-Dimensional Quantum Computing}

\author{Kevin Mato}%
 \email{kevin.mato@tum.de}
\affiliation{%
 Chair for Design Automation, Technical University of Munich, Munich, Germany
}%
\author{Martin Ringbauer}%
 \email{martin.ringbauer@uibk.ac.at}
\affiliation{%
 Institut für Experimentalphysik, University of Innsbruck, Innsbruck, Austria
}%
\author{Lukas Burgholzer}
 \email{lukas.burgholzer@tum.de}
\affiliation{
 Chair for Design Automation, Technical University of Munich, Munich, Germany
}%
\author{Robert Wille}
 \email{robert.wille@tum.de}
\affiliation{
 Chair for Design Automation, Technical University of Munich, Munich, Germany
}%
\affiliation{
Software Competence Center Hagenberg (SCCH) GmbH, Hagenberg, Austria
}

\begin{abstract}
Quantum computing holds great promise for surpassing the limits of classical devices in many fields. Despite impressive developments, however, current research is primarily focused on qubits. At the same time, quantum hardware based on multi-level, qudit, systems offers a range of advantages, including expanded gate sets, higher information density, and improved computational efficiency, which might play a key role in overcoming not only the limitations of classical machines but also of current qubit-based quantum devices.
However, working with qudits faces challenges not only in experimental control but particularly in algorithm development and quantum software. In this work, we introduce \emph{MQT Qudits}, an open-source tool, which, as part of the \emph{Munich Quantum Toolkit} (MQT), is built to assist in designing and implementing applications for mixed-dimensional qudit devices. We specify a standardized language for mixed-dimension systems and discuss circuit specification, compilation to hardware gate sets, efficient circuit simulation, and open challenges. MQT Qudits is available at \url{github.com/cda-tum/mqt-qudits} and on pypi at \url{pypi.org/project/mqt.qudits/}.
\end{abstract}

\maketitle

\section{Introduction} 
Quantum computing is an emerging information processing paradigm that promises to solve certain industrial and scientific problems up to exponentially faster than classical devices. This potential as spurred impressive developments in the field of quantum computing over the past decades. State-of-the-art devices now host hundreds of noisy \emph{quantum bits}~(qubits) and support a limited number of logical operations on these qubits. Such devices are referred to as \emph{Noisy Intermediate-Scale Quantum}~(NISQ) devices~\cite{preskill2018quantum} and have been realized in a number of technology platforms, including superconducting circuits~\cite{Arute2019}, trapped ions~\cite{Pogorelov2021}, neutral atoms~\cite{Bluvstein_2023}, and single photons~\cite{Flamini2018a}. 
Notably, these devices almost exclusively work with two-level qubits, while the underlying hardware almost always natively supports encoding multi-valued logic in high-dimensional \emph{quantum digits}~(qudits)~\cite{ringbauer2021universal, Morvan2020, Chi2022}.

Research on qudit design and computation has a long history, with efforts having primarily focused on \mbox{proof-of-principle} experimental control, conceptual studies of primitives~\cite{Lanyon2008} and comparisons of algorithms for idealized qudits and qubits~\cite{Wang2020}. Fundamentally, a qudit can not only store and process more information per quantum particle, but also feature a richer Hilbert space structure~\cite{Ringbauer2017, kraft2018} and, thus, a larger set of logical operations that make processing more efficient~\cite{Gao2023}. \mbox{Proof-of-principle} experimental demonstrations~\mbox{\cite{Lanyon2008, Gedik2015, Zhan2015, meth2024simulating2dlatticegauge}} have shown that these qudit advantages translate into improvements in circuit complexity and algorithmic efficiency for a wide range of problems. 
%
More recently, efforts have intensified with the demonstration of universal qudit quantum processors with competitive performance~\cite{ringbauer2021universal, Morvan2020, Chi2022}. Such devices are already being used in quantum simulation applications, where a mixed-dimensional approach is much more naturally suited~\cite{meth2024simulating2dlatticegauge, edmunds2024, Navickas2024}. Examples include the simulation of high-spin systems~\cite{Haldane1983, Senko2015, edmunds2024}, quantum chemistry~\cite{Navickas2024}, and lattice gauge theories~\cite{Cuadra2022, meth2024simulating2dlatticegauge, Calajo2024, Popov2024}.
%
Combining the improved efficiency from native, rather than qubit-transpiled, implementations with the improved gate complexity enabled by the temporary expansion of the Hilbert space, has the potential to enable significantly shorter quantum circuits in mixed-dimensional hardware. Yet optimizing the use of mixed-dimensional resources is a highly non-trivial challenge.

In this work, we introduce MQT Qudits, an open-source framework for mixed-dimensional quantum computing, which is part of the \emph{Munich Quantum Toolkit}~(MQT, \cite{wille2024mqthandbooksummarydesign}). MQT Qudits provides efficient, automated, and accessible methods for tackling challenging problems in the design of quantum computing applications. In the following, we will provide a brief overview of MQT Qudits from both a user's perspective (showcasing how to utilize the tool for describing problems and algorithms in the form of quantum circuits, and how to simulate and compile them) and a technical perspective (covering the underlying methods, the choices made during development, and how to contribute to the framework). More precisely, we will:
\begin{itemize}
\item first, review the background and walk through the advantages, challenges, and opportunities of mixed-dimensional quantum computing (Section~\ref{sec:Back}),
\item illustrate how to get started with MQT Qudits (Section~\ref{sec:Get}),
\item introduce the languages and interfaces used by the tool (Section~\ref{sec:Lang}), and
\item present the corresponding methods for quantum circuit simulation (Section~\ref{sec:Sim}) and quantum circuit compilation (Section~\ref{sec:Comp}).
\end{itemize}


\section{Background} \label{sec:Back}
This section gives a brief overview of the principles of qudit quantum computing in current hardware before going into MQT Qudits. It should be mentioned that a thorough examination of the vast topic of quantum computation is not possible within this brief study. The interested reader is referred to in-depth literature, such as Ref.~\cite{Wang2020}.

\subsection{Quantum Information Processing}
In the following, we will consider the qudit (quantum digit) as the fundamental unit of information. In contrast to the binary counterpart, the state of a qudit is represented by a vector in a $d$-dimensional Hilbert space, $\mathcal{H}_d$. This state can be expressed as a linear superposition:

\begin{equation}
\ket{\psi} = \alpha_0\ket{0} + \alpha_1\ket{1} + \cdots + \alpha_{d-1}\ket{d-1},
\end{equation}

where the set $\{\ket{i}\}$ is referred to as the computational basis, and $|\alpha_i|^2$ is the probability of observing the state $\ket{i}$ upon measurement, with $\sum_{i=0}^{d-1} |\alpha_i|^2 = 1$. Qudits offer a richer state space, enabling more complex quantum information processing (QIP) than their binary counterpart, which are typically obtained by restricting naturally multi-level systems to just two levels. As a result, the core properties of quantum systems, superposition and entanglement, become much more interesting.


\begin{example}
\label{ex:state}
Consider a single qudit with three states (also referred to as \emph{qutrit}).
The quantum state $\ket{\psi} = \sqrt{\frac{1}{3}}\ket{0} + \sqrt{\frac{1}{3}}\ket{1} + \sqrt{\frac{1}{3}}\ket{2}$ corresponds to an equal superposition of the three basis states. This state can equivalently be represented as the vector $\sqrt{\frac{1}{3}} \begin{bmatrix} 1, 1, 1 \end{bmatrix}^T$.

Similarly, consider a composite system of one qutrit and one qubit in the (mixed-dimensional) entangled state \mbox{$\ket{\psi'} = \sqrt{\frac{1}{3}}\ket{0}_3\ket{0}_2 + \sqrt{\frac{1}{3}}\ket{1}_3\ket{1}_2 + \sqrt{\frac{1}{3}}\ket{2}_3\ket{0}_2$}---equivalently represented by the vector $\sqrt{\frac{1}{3}} \begin{bmatrix} 1 , 0 , 0 , 1 , 1 , 0 \end{bmatrix}^T$.
\end{example}

The manipulation of qudits is achieved through unitary operations, represented by $d \times d$ matrices $U\in \text{SU}(d)$. In mixed-dimensional architectures, comprising qudits of varying dimensions, the unitary matrix governing a multi-qudit circuit scales as $\prod_i d_i \times \prod_i d_i$, where $d_i$ denotes the dimension of each qudit.

\begin{example}
    A simple example of qudit operations are the generalizations of the Pauli X and Z~\cite{Clark_2006}. In the case of a qutrit, they are described as
\begin{equation}
\label{eq:pauli}
        X = 
        \begin{bmatrix}
            0 & 0 & 1 \\
            1 & 0 & 0 \\
            0 & 1 & 0 \\
        \end{bmatrix} \quad
        Z = 
        \begin{bmatrix}
            1 & 0 & 0 \\
            0 & \omega & 0\\
            0 & 0 & \omega^2\\
        \end{bmatrix},
\end{equation}
with $\omega = e^{\frac{2\pi i}{d}}$ and $d$ being the dimension of the single qudit. 
\end{example}

Single-qudit operations can be tailored to act on specific subspaces or the entire Hilbert space of the qudit. Subspace operations could, for example, involve rotations built from Gell-Mann matrices~\cite{ringbauer2021universal}, which physically describe two-level couplings in subspaces of the qudit. Operations on the full Hilbert space frequently utilize rotations built from the Pauli generalizations~\cite{Clark_2006}, called also clock and shift operators. Entangling operations can be engineered by composing Hamiltonians using these fundamental operators. The power of quantum algorithms lies in the efficient manipulation of superpositions of quantum states many-body interference effects and entanglement. Through careful sequencing of quantum operations, we can construct circuits that exploit this interference, and potentially offer computational advantages over classical algorithms for certain problems. 



\subsection{Advantages, Challenges and Opportunities}



After discussing the fundamental aspects of quantum computing and its implementation using \mbox{mixed-dimensional} quantum systems, we now turn our attention to the advantages, challenges, and opportunities presented by \mbox{mixed-dimensional} quantum computing. 

Current qubit-based quantum computers offer potential advantages in solving highly complex problems by reducing the amount of computational resources, yet scalability remains challenging. Qudit-based quantum computers could mitigate these limitations to some extent by increasing the information density and computational efficiency for a given number of quantum information carriers.


For instance, qudits enable native encoding of \mbox{mixed-dimensional} problems, such as those often found in quantum simulation tasks. Rather than requiring $\lceil{\log(d)}\rceil $ qubits to encode a $d$-dimensional object, thereby turning simple two-body into complicated many-body interactions, the mixed-dimensional approach enables us to allocate only the required resources and maintain simple circuit structures. In some cases, this may even allow for the flexible truncation of certain quantities to seamlessly enable the computation at different levels of accuracy~\cite{meth2024simulating2dlatticegauge}. However, these varying degrees of freedom become part of the problem and have to be tracked. Being able to describe the problems easily and efficiently is the first step towards the exploitation of emerging \mbox{mixed-dimensional} quantum processors.


The introduction of qudits in quantum computing architectures creates a lot more freedom for design choices and problem-tailored solutions. This includes not only the choice and evolution of the single-qudit encoding but also the wide range of possible local and entangling operations. While for qubits the controlled-NOT (CNOT) gate is the central resource, qudit systems can build on an ever-increasing collection of non-equivalent entangling gates with varying degrees of entangling power~\cite{Huber2013} to enable efficient circuit design.
%
%
The flip side of this greatly increased freedom and flexibility is the increased complexity of designing short and efficient, noise-resilient quantum circuits for qudit systems. Finding a way through this landscape of advantages, challenges, and opportunities presented by mixed-dimensional quantum systems necessitates a comprehensive set of methodologies and automated software tools for effective exploration and utilization, such as classical simulators and compilers.


To this end, the MQT Qudits framework offers an integrated advanced language for quantum circuit representations, simulation, and compilation methods, and software that facilitates the utilization of \mbox{mixed-dimensional} quantum systems. This framework enables researchers to systematically exploit the potential for quantum information processing tasks across diverse technology platforms and to access the full potential of \mbox{mixed-dimensional} qudit systems.

The remainder of the paper will introduce each component of the framework from two different perspectives: one focusing on the functionality for users, and the other on the technical aspects for motivated scientists and developers. We will always begin with the former.

\section{Getting Started} \label{sec:Get}
MQT Qudits empowers users to efficiently create, simulate, compile, and enhance mixed-dimensional quantum circuits across diverse quantum computing platforms. In the next sections, we provide a concise overview of how to set up and begin using MQT Qudits.
The library is primarily written in Python, with components such as simulation engines written in C++ for performance reasons and then interfaced with Python. The installation process is designed to be user-friendly and does not require explicit compilation of the source. MQT Qudits supports Windows, Linux, and Mac for all versions of Python from 3.9 to 3.12. Installing MQT Qudits simply requires the execution of the following lines:

\begin{minted}[fontsize=\footnotesize, frame=lines, xleftmargin=10pt , escapeinside=||]{bash}
$ python3 -m venv .venv
$ source .venv/bin/activate
(.venv) $ pip install mqt.qudits
\end{minted}

The code above demonstrates how to create a new environment and subsequently install the library directly from the terminal.
After setting up the working environment, it is possible to write your first program.

\section{Languages and Interfaces}\label{sec:Lang}
The first step towards simulating and compiling mixed-dimensional quantum circuits, which could potentially solve complex problems, is to define a language that can represent them. None of the existing languages provide appropriate means for this purpose. 

\subsection{The User's Perspective}
MQT Qudits introduces DITQASM, an adaptation of OpenQASM~\cite{cross2017openquantumassemblylanguage} for describing mixed-dimensional quantum circuits and architectures. QASM languages serve as a bridge between high-level quantum algorithms and their physical implementation on hardware, and in this case, DITQASM introduces new fundamental features, as illustrated in Figure \ref{fig:qasm_example}. Users can use DITQASM to translate abstract quantum algorithms into executable forms, analyze circuit complexity, and study the impact of noise and error correction in real quantum systems.
Key features include:
\begin{itemize}
    \item Register allocation: descriptions of the quantum and classical registers involved and their dimensionality.
    \item Gate-level description: the precise specification of quantum gates and their application to qudits of different dimensions, with a new way of controlling gates on arbitrary levels.
    \item Hardware-agnostic representation: a standardized way to express quantum circuits across different quantum computing platforms.
    \item Customizability and extendibility.
\end{itemize}

\begin{example}
Figure~\ref{fig:qasm_example} shows the DITQASM implementation of a quantum program using two qudit registers: one containing two qudits with dimensions 2 and 3, and another containing two qudits with dimensions 4 and 7 (lines 2-3). Additionally, a classical register is declared with 4 units (line 4). The circuit is then constructed, beginning with a Hadamard gate (\tquote{h}) applied on the first qudit of register 2 and controlled by the state of the two qudits in the other register. This control is described using the \tquote{ctl} syntax (line 5), a new feature introduced in DITQASM. This is followed by a controlled-sum gate \tquote{csum}~\cite{Mato23_Ent_Comp} (line 6) and two subspace rotations \tquote{rxy}~\cite{ringbauer2021universal}. The definitions of the gates do not require any knowledge of the technology platform implementing the multi-level quantum systems, and the dimensionality is not required at the application of the gates, but the focus is only on the semantic information of the rotations. The measurement is performed according to the default options (lines 9-12).
\end{example}

Alternatively, circuits can be described by a Python interface, as shown in Figure \ref{fig:python_language}. Here, the Python program starts with the import of the two main components of a quantum circuit: quantum registers, a quantum circuit, and, optionally, classical registers (lines 1-3). After creating a quantum circuit instance, the next two lines allocate two qudit registers, and a classic register (lines 5-11): ``\texttt{reg\_1}'' and another one named ``\texttt{reg\_2}.'' The ability to name and allocate different registers facilitates the direct translation of interactions, such as those represented on a lattice or a graph, into an algorithm. 
Subsequently, the quantum operations are applied: a Hadamard gate on the ``\texttt{reg\_2}'' register and a controlled-sum gate, which generalizes the CNOT gate (lines 12-14). The interpreter tracks the dimensionality of the qudits and applies the corresponding operations. The program concludes with a measurement of the qudits.

The Python interface and DITQASM also support various gates, including hardware-specific ones like the Molmer-Sorenson~(MS) gate from ~\cite{ringbauer2021universal} and the generalized light-shift~(LS), a genuine qudit entangling gate~\cite{Sawyer2021,nativequdit}, as well as custom unitary evolutions on one, two, or multiple qudits and qubits.

\begin{figure}[t]
\begin{minted}[fontsize=\footnotesize, frame=lines, linenos, xleftmargin=10pt , escapeinside=||]{text}
DITQASM 2.0;
qreg reg_1 [2][2, 3];
qreg reg_2 [2][4, 7];
creg meas[4];
h reg_2[0] ctl reg_1[0] reg_1[1] [0,0];
csum reg_2[0], reg_1[0];
rxy (0, 2, pi, pi/2) reg_1[1]; 
rxy (0, 1, pi, pi/2) reg_2[1]; 
measure reg_1[0] -> meas[0];
measure reg_1[1] -> meas[1];
measure reg_2[0] -> meas[2];
measure reg_2[1] -> meas[3];
\end{minted}
\caption{Example of DITQASM representing a program run on a quantum architecture made of two registers, with qudits of dimensions 2, 3, 4, and 7.}
\label{fig:qasm_example}
\end{figure}

\begin{small}
\begin{figure}[t]
\begin{minted}[fontsize=\footnotesize, frame=lines, linenos, xleftmargin=10pt ]{python}
from mqt.qudits.quantum_circuit import QuantumCircuit
from mqt.qudits.quantum_circuit import QuantumRegister,\
                                        ClassicRegister

circuit = QuantumCircuit()
reg_1 = QuantumRegister("reg_1", 2, [2, 3])
reg_2 = QuantumRegister("reg_2", 2, [4, 7])
meas = ClassicRegister("meas", 4)
circuit.append(reg_qudit)
circuit.append(qubit_reg)
circuit.append_classic(meas)
circuit.h(reg_2[0]).control([reg_1[0],\
                                reg_1[1]], [0, 0])
circuit.csum([reg_2[0], reg_1[0]])
circuit.r(reg_1[1], [0, 2, np.pi, np.pi/2])
circuit.r(reg_2[1], [0, 1, np.pi, np.pi/2])
\end{minted}
\caption{A simple quantum program written for a mixed-dimensional quantum computer, in Python, through MQT Qudits.}
\label{fig:python_language}
\end{figure}
\end{small}

\subsection{The Technical Perspective}
The section gives an intuition of the functioning of the DITQASM interpreter and pointers for further extensions and improvements. The distinguishing feature of DITQASM is its capacity to define arbitrary Hilbert space dimensions for individual elements within a quantum register, a capability previously unavailable. The expanded gate set, equipped with automatic dimensionality matching, makes the language hardware-agnostic. Furthermore, the automatic generalization of single-qudit and entangling gates to higher dimensions abstracts away the internal structure and properties of individual qudits, facilitating a focus on algorithm development. The language supports fine-grained control over multi-level systems, allowing for the specification of arbitrary control levels within the range $[0, d_i-1]$ for the control qudit in controlled operations. Classical registers are implemented as integer wrappers, maintaining consistency with the multi-level quantum paradigm.

The current Python 3-based DITQASM interpreter is in its nascent stages. The current version (0.1.0) employs pattern matching during the parsing of the text of the quantum program; the instructions will be collected, and all the information and metadata will be tracked in the software. This data is then used to generate all the corresponding components and eventually compose them in the correct fashion. This approach ensures extensibility, allowing for easy incorporation of additional gates and features as the field evolves. 

The Python interface and DITQASM are in direct relation to each other. This is sufficient for a local prototyping environment but will not suffice for large-scale deployments or for using MQT Qudits in a quantum computer-centric HPC environment. Possible solutions would involve the development of a suitable \emph{Quantum Intermediate Representation}~(QIR) for mixed-dimensional systems. Prospective developments include the integration of classical control flow syntax and mid-circuit measurements, crucial for implementing error correction schemes in qudit systems. These advancements are particularly pertinent given the growing relevance of qudit-based approaches across various domains of quantum computing research.

For a more extensive overview of all available options, techniques, and advanced features, please refer to MQT Qudits documentation at \url{mqt.readthedocs.io/projects/qudits/en/latest/}. This documentation provides in-depth information on working with \mbox{mixed-dimensional} quantum circuits, including detailed API references, tutorials and examples, as well as advanced configuration options.

\section{Quantum Circuit Simulation}\label{sec:Sim}
Quantum circuit simulators are essential in assessing the quality of quantum circuits and quickly iterating new design choices for the algorithms to run. They enable the exploration of alternative qudit-based approaches, circuit optimizations, and the development or error mitigation schemes without having to rely on physical hardware. Different types of quantum circuits and applications require specialized simulators for each task. To this end, MQT Qudits offers two simulation options, each one with its merits and limitations. Again, we will examine both from the user and technical perspectives.

\subsection{The User's Perspective}
MQT Qudits offers a user-friendly interface that hides the complexity of simulating quantum circuits.
In Figure~\ref{fig:basic_simulation}, a simple demonstration of how to use the framework for quantum circuit simulations is shown. There are two levels of complexity for performing a simulation depending on the required level of customization. Once a quantum circuit is defined, you can call the simulate method of the circuit instance, which will return a job object upon completion of the execution. It is possible to retrieve the quantum state from the job, by calling \tquote{\texttt{get\_state\_vector()}}. Alternatively, it is possible to choose a backend, thanks to the method \tquote{\texttt{get\_backend(backend\_name)}}, between a tensor network simulator (\tquote{tnsim}), a mixed-dimensional decision diagram simulator (\tquote{misim},~\cite{MatoDD}), and a suite of simple emulators representing physical devices, again by using \tquote{\texttt{get\_backend(fake\_backend\_name)}}. These latter, prefixed with \tquote{fake}, facilitate compilation and noisy simulation of quantum circuits with device-specific properties. Upon backend selection, the environment is ready to run the simulation of the circuit. 

A particularly relevant feature of a quantum circuit simulator is noise-aware simulation to allow for estimating the quality and feasibility of execution on a given quantum hardware. Figure~\ref{fig:noise_simulation} shows how to construct a noise model by specifying probabilities for generalized Pauli errors, analogous to X and Z flips in the qubit case, extended to the qudit Hilbert space and subspaces thereof. The framework allows for fine-grained control over noise channels, supporting noise on local, as well as, multi-qudit operations with different effects on target and control qudits, and various other noise channels as detailed in the documentation. This approach enables the simulation of realistic noise processes in higher-dimensional quantum systems, crucial for assessing the performance of qudit-based quantum algorithms and error correction schemes. Different types of noise instances can be combined in a noise model and input as parameters of a simulation. The knowledge about the noise of a system can be easily integrated by adding it as an option to the simulation or by choosing a \tquote{fake} backend that will possess this knowledge by default. 

\begin{small}
\begin{figure}[t]
\begin{minted}[fontsize=\footnotesize, frame=lines, linenos, xleftmargin=10pt ]{python}
# import components and circuit
from mqt.qudits.simulation import MQTQuditProvider

# program written before
state = circuit.simulate()

# alternatively
provider = MQTQuditProvider()
backend = provider.get_backend("tnsim")
job = backend.run(circuit)
result = job.result()
state = result.get_state_vector()
\end{minted}
\caption{Simulation of a qudit circuit with MQT Qudits' TnSim.}
\label{fig:basic_simulation}
\end{figure}
\end{small}

\begin{small}
\begin{figure}[t]
\begin{minted}[fontsize=\footnotesize, frame=lines, linenos, xleftmargin=10pt ]{python}
# import components and circuit
from mqt.qudits.simulation import MQTQuditProvider
from mqt.qudits.simulation.noise_tools.noise import \
    Noise, NoiseModel

local_error = Noise(0.01, 0.001)
local_error_rz = Noise(0.018, 0.002)

noise_model = NoiseModel() 
noise_model.add_quantum_error_locally(local_error,\
                        ["h", "rxy", "s", "x", "z"])
noise_model.add_quantum_error_locally(local_error_rz,\
                        ["rz"])

provider = MQTQuditProvider()
backend = provider.get_backend("tnsim")
job = backend.run(circuit, noise_model=noise_model)
result = job.result()
counts = result.get_counts()
\end{minted}
\caption{Simulation of a qudit circuit with a tailored noise model through MQT Qudits.}
\label{fig:noise_simulation}
\end{figure}
\end{small}

\subsection{The Technical Perspective}
The simplicity of use of the simulation components of MQT Qudits is due to the implementation of a modular architecture. The entry point is a provider object that serves as an abstraction layer for various backend implementations. For brevity, this section will cover only the most novel backends of the library: MiSiM~\cite{MatoDD} and TnSim. MiSiM, a C++ implementation, extends \mbox{edge-weighted} \emph{Decision Diagrams} (DDs,~\cite{ZulehnerW19, complex_ZulehnerHW19}), or QMDDs~\cite{NiemannWMTD16}, to efficiently simulate mixed-dimensional qudit systems. By dynamically capturing qudit dimensionality, MiSiM significantly reduces memory requirements, enabling the simulation of complex circuits, e.g., exceeding, in some cases, 100 qutrits. This DD-based approach exploits redundancies and symmetries in the representation of quantum states and operations, allowing direct manipulation of operations on the diagram structure. 

Additionally, DDs can efficiently represent sparse data, becoming a powerful tool for simulating circuits with \mbox{non-trivial} quantum correlations by representing states that would otherwise require exponential resources. TnSim, conversely, utilizes \emph{tensor networks} representations via Google's TensorNetwork API~\cite{roberts2019tensornetwork} in Python. The tensor network simulator contained in MQT Qudit takes care of the construction of the network and the dynamical allocation of nodes in it and leverages optimized tensor contraction algorithms, offering performance advantages for circuits with specific entanglement structures. Tensor networks efficiently represent quantum many-body states with limited entanglement, particularly those obeying area-law scaling~\cite{Or_s_2014}. They decompose high-dimensional states into contracted lower-dimensional tensors, reducing computational complexity from exponential to polynomial for many physical systems. 

Both backends support noise-aware simulation of quantum circuits through a stochastic approach~\cite{Berquist, GrurlKFW21, noise_aware_grurl}. This method appends qudit Pauli X or Z gates after each operation with a specific probability described in the noise instance (Figure~\ref{fig:noise_simulation}). For gates operating in subspaces, such as local subspace rotations~\cite{ringbauer2021universal}, the noise is applied as subspace X or Z operations. The noise model in Fig.~\ref{fig:noise_simulation} encodes the logic determining whether each gate has a noise operation appended, and for entangling operations, whether noise is applied to the target, control(s), or a subset of qudits involved. A particularly simple, yet highly relevant, extension is the integration of numerical distortions of the gate parameters in the circuit associated with a specific noise channel, or the inclusion of subspace error operations for all types of gates.

This stochastic noise simulation method efficiently models decoherence and gate errors in quantum systems without the need for full density-matrix calculations. However, it requires multiple simulation runs to create a statistical distribution of results, which fluctuates due to operation-specific noise. Parallel processing could conceivably mitigate this limitation by executing runs simultaneously on several processes. 

In summary, the library provides a suite of complementary simulators that leverage two data structures and different implementations. These can provide solutions with varying performance depending on the use case, the complexity of the noise to simulate, and the user's preference. Users can choose the most suitable simulator based on their specific requirements and computational constraints.

\section{Quantum Circuit Compilation}\label{sec:Comp}
Qudit compilation transforms abstract quantum algorithms into executable sequences for specific hardware, bridging theory, and implementation. The MQT Qudits compiler integrates quantum control theory, optimization, and custom heuristics to address coherence and fidelity challenges. It encompasses gate decomposition, mapping, and noise adaptation, crucial for realizing quantum algorithmic potential in near-term qudit processors. The compiler aims to be accessible to a wide range of users, from novices exploring mixed-dimensional quantum computing to well-resourced laboratories, while offering flexible, modular, and computationally efficient software that balances user-friendly operation with advanced technical capabilities. This section illustrates the use of the compiler.

\subsection{The User Perspective} 
Figure~\ref{fig:compiler} outlines how the user can exploit MQT Qudits for compilation tasks.
The code shows how to set the initial state of the quantum circuit as a first operation, starting from the numpy array~\cite{harris2020array} of the state. After setting the initial state (line 8), the user can define the rest of the quantum circuit and directly compile the circuit for a target device (lines 9-11), just by stating the name of the device, prefixed by \tquote{fake}. The compilation, in this case, chooses some default flags and returns the compiled circuit compatible with the device to be executed on. Although highly flexible, the compiler is constantly evolving. Multi-body gate compilation will be supported in future releases.
For the user who is interested in benchmarking and tuning the compilation process (lines 14-20), the compilation of a quantum circuit requires minimal code and only three steps, namely 
\begin{enumerate}
    \item instantiation of the qudit compiler, 
    \item selection of compilation steps, and
    \item choice of a target backend.  
\end{enumerate}

In this case, the compiler is returning the compiled circuit, which can then be either simulated or passed to the API of the actual device, after using the correct backend, e.g., \tquote{innsbruck01}.

\begin{small}
\begin{figure}[t]
\begin{minted}[frame=lines, linenos, xleftmargin=10pt ]{python}
# import components and circuit
from mqt.qudits.simulation import MQTQuditProvider
from mqt.qudits.compiler import QuditCompiler
provider = MQTQuditProvider()

# optionally the user can set the initial state
state = np.array(...)
circuit.set_initial_state(state)
# rest of the program
# simple interface
new_circuit = circuit.compile("faketraps2six")

# alternatively
qudit_compiler = QuditCompiler()
# choice of the passes
passes = ["PhyLocQRPass", "PhyEntQRPass"]

device = provider.get_backend("faketraps2six")
new_circuit = qudit_compiler.compile(device, \
                        circuit, passes)
\end{minted}
\caption{A prototypical compilation of a quantum circuit on MQT Qudits.}
\label{fig:compiler}
\end{figure}
\end{small}

\subsection{The Technical Perspective}
Shielding users from the complexity of the compilation process necessitates the development of sophisticated methods and implementations. The compiler is designed as a modular, pass-based tool that prioritizes algorithmic efficiency, despite being entirely written in Python. In the following, we briefly review the main methods currently implemented in the MQT Qudits compiler, namely a state preparation routine, a single-qudit operations compiler, and a two-qudit operations compiler.
The state preparation routine takes a quantum state in the form of a numpy array and outputs a sequence of local and controlled two-level rotation, with an increasing number of controls. The state preparation makes use of mixed-dimensional decision diagrams~\cite{MatoDD} implemented in Python for efficiently representing the state. A synthesis algorithm reads the DD structure, possibly approximates it if required by the user, and returns a sequence of quantum operations to generate the state. The execution of the multi-controlled operations is allowed by an implicit decomposition in qudit \tquote{CNOT} by using ancilla levels in the qudits~\cite{meth2024simulating2dlatticegauge}. We refer to the paper~\cite{Mato24_State_Prep} for further details on the synthesis.

The compilation steps enabled in the process depend on the user's choice of passes. Depending on the selected passes, the compilation can target either \mbox{single-qudit} or \mbox{two-qudit} unitaries. The pass nomenclature indicates whether the compilation produces a physically executable sequence compatible with the backend (\tquote{Phys}) or a logical sequence (\tquote{Log}). \tquote{Log} denotes single-qudit unitary compilation, while \tquote{Ent} signifies two-qudit unitary compilation.

\begin{figure}
    \centering
    \includesvg[width=0.3\textwidth]{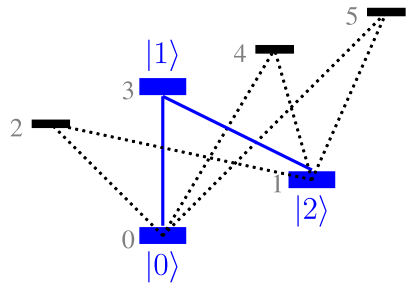}
    \caption{Energy level graph of a qutrit, 3-level qudit.}
    \label{fig:e_graph}
\end{figure}

Single-qudit unitaries are compiled using sequences of 2-level subspace rotations. This reconstruction method aligns with the energy level graph of the individual qudit, as shown schematically in Figure~\ref{fig:e_graph}. The energy level graph~\cite{Mato22_Single} is a data structure that represents the internal structure of the qudit, storing information about the quality of rotations between two levels and tracking various properties of the decomposition on the physical system. This graph structure corresponds to the coupling map between the various qudit levels. The use of the energy level graph and graph optimization heuristics has led to improvements in the compilation of single-qudit unitaries, particularly in reducing the number of rotations required, especially phase rotations~\cite{Mato22_Single}.

MQT Qudits compiles two-qudit unitaries of arbitrary dimensionality into sequences of controlled rotations (crot) and partial swap operations (pswap). These operations are then either directly implemented in hardware or further decomposed into a sequence of local operations and controlled-exchange gates, which is a qudit-embedded 2-level CNOT gate. This methodology~\cite{Mato23_Ent_Comp} provides a computationally efficient means of compiling arbitrary interactions into standardized entangling gates, independent of the underlying hardware architecture. The compiler allows for variational compilation of the standardized entangling gates into the hardware-native gate set, tailoring the circuit to the specific hardware. Further extensions should include tools for compressing the generated circuits to further increase the efficiency.

An additional standalone tool is the qubit compression mapper~\cite{Mato23_Compr}, a utility that recommends dimensionality adjustments in mixed-dimensional architectures to efficiently encode qubit circuits into qudit circuits. This tool facilitates the translation between different quantum information encoding schemes, potentially enabling more efficient use of available quantum resources.

\section{Conclusion}
\label{sec:conclusion}

MQT Qudits provides an open-source framework for mixed-dimensional quantum computing, including flexible automated tools for the design, specification, simulation, and compilation of mixed-dimensional quantum circuits. Considering the fundamental differences between qubit and qudit QIP, the framework is designed to be easily extendable to new kinds of gate operations that are expected to be developed as the field evolves. Similarly, the development and inclusion of more hardware-specific noise models for the various types of qudit quantum computing hardware is a point of great interest to enable accurate circuit simulation. 

A particularly interesting topic is the dynamic evolution of the data structures during the simulation with the goal of tracking the usage of ancilla levels in the qudit. Finally, the compiler framework enables flexible and adaptive compilation, yet does not always achieve optimal circuits. Future efforts should prioritize automated multi-qudit gate compilation, and optimizing two-qudit gate compilations to minimize the resulting circuit complexity. Efforts should also include resynthesis procedures to reduce the final circuit depth. In order to facilitate these developments, MQT Qudits will, in the future, include scalable verification methods for \mbox{mixed-dimensional} systems.

\begin{acknowledgments}
{\small
We thank Stefan Hillmich for his valuable insights and support throughout the early development of the core ideas in this work. This research was funded by the European Union under the Horizon Europe Programme as part of the project NeQST (Grant Agreement 101080086) as well as by the European Research Council projects DA QC (Grant Agreement 101001318) and QUDITS (Grant Agreement 101080086). Views and opinions expressed are however those of the author(s) only and do not necessarily reflect those of the European Union or the European Research Council Executive Agency. Neither the European Union nor the granting authority can be held responsible for them. 
The work is part of the Munich Quantum Valley, which is supported by the Bavarian state government with funds from the Hightech Agenda Bayern Plus and was partially supported by the BMK, BMDW, and the State of Upper Austria in the frame of the COMET program (managed by the FFG).}
\end{acknowledgments}


\bibliography{apssamp}

\begin{thebibliography}{43}%
\makeatletter
\providecommand \@ifxundefined [1]{%
 \@ifx{#1\undefined}
}%
\providecommand \@ifnum [1]{%
 \ifnum #1\expandafter \@firstoftwo
 \else \expandafter \@secondoftwo
 \fi
}%
\providecommand \@ifx [1]{%
 \ifx #1\expandafter \@firstoftwo
 \else \expandafter \@secondoftwo
 \fi
}%
\providecommand \natexlab [1]{#1}%
\providecommand \enquote  [1]{``#1''}%
\providecommand \bibnamefont  [1]{#1}%
\providecommand \bibfnamefont [1]{#1}%
\providecommand \citenamefont [1]{#1}%
\providecommand \href@noop [0]{\@secondoftwo}%
\providecommand \href [0]{\begingroup \@sanitize@url \@href}%
\providecommand \@href[1]{\@@startlink{#1}\@@href}%
\providecommand \@@href[1]{\endgroup#1\@@endlink}%
\providecommand \@sanitize@url [0]{\catcode `\\12\catcode `\$12\catcode `\&12\catcode `\#12\catcode `\^12\catcode `\_12\catcode `\%12\relax}%
\providecommand \@@startlink[1]{}%
\providecommand \@@endlink[0]{}%
\providecommand \url  [0]{\begingroup\@sanitize@url \@url }%
\providecommand \@url [1]{\endgroup\@href {#1}{\urlprefix }}%
\providecommand \urlprefix  [0]{URL }%
\providecommand \Eprint [0]{\href }%
\providecommand \doibase [0]{https://doi.org/}%
\providecommand \selectlanguage [0]{\@gobble}%
\providecommand \bibinfo  [0]{\@secondoftwo}%
\providecommand \bibfield  [0]{\@secondoftwo}%
\providecommand \translation [1]{[#1]}%
\providecommand \BibitemOpen [0]{}%
\providecommand \bibitemStop [0]{}%
\providecommand \bibitemNoStop [0]{.\EOS\space}%
\providecommand \EOS [0]{\spacefactor3000\relax}%
\providecommand \BibitemShut  [1]{\csname bibitem#1\endcsname}%
\let\auto@bib@innerbib\@empty
\bibitem [{\citenamefont {Preskill}(2018)}]{preskill2018quantum}%
  \BibitemOpen
  \bibfield  {author} {\bibinfo {author} {\bibfnamefont {J.}~\bibnamefont {Preskill}},\ }\bibfield  {title} {\bibinfo {title} {Quantum computing in the {NISQ} era and beyond},\ }\href@noop {} {\bibfield  {journal} {\bibinfo  {journal} {Quantum}\ }\textbf {\bibinfo {volume} {2}},\ \bibinfo {pages} {79} (\bibinfo {year} {2018})}\BibitemShut {NoStop}%
\bibitem [{\citenamefont {Arute}\ \emph {et~al.}(2019)\citenamefont {Arute}, \citenamefont {Arya}, \citenamefont {Babbush}, \citenamefont {Bacon}, \citenamefont {Bardin}, \citenamefont {Barends}, \citenamefont {Biswas}, \citenamefont {Boixo}, \citenamefont {Brandao}, \citenamefont {Buell}, \citenamefont {Burkett}, \citenamefont {Chen}, \citenamefont {Chen}, \citenamefont {Chiaro}, \citenamefont {Collins}, \citenamefont {Courtney}, \citenamefont {Dunsworth}, \citenamefont {Farhi}, \citenamefont {Foxen}, \citenamefont {Fowler}, \citenamefont {Gidney}, \citenamefont {Giustina}, \citenamefont {Graff}, \citenamefont {Guerin}, \citenamefont {Habegger}, \citenamefont {Harrigan}, \citenamefont {Hartmann}, \citenamefont {Ho}, \citenamefont {Hoffmann}, \citenamefont {Huang}, \citenamefont {Humble}, \citenamefont {Isakov}, \citenamefont {Jeffrey}, \citenamefont {Jiang}, \citenamefont {Kafri}, \citenamefont {Kechedzhi}, \citenamefont {Kelly}, \citenamefont {Klimov}, \citenamefont {Knysh}, \citenamefont {Korotkov},
  \citenamefont {Kostritsa}, \citenamefont {Landhuis}, \citenamefont {Lindmark}, \citenamefont {Lucero}, \citenamefont {Lyakh}, \citenamefont {Mandr{\`{a}}}, \citenamefont {McClean}, \citenamefont {McEwen}, \citenamefont {Megrant}, \citenamefont {Mi}, \citenamefont {Michielsen}, \citenamefont {Mohseni}, \citenamefont {Mutus}, \citenamefont {Naaman}, \citenamefont {Neeley}, \citenamefont {Neill}, \citenamefont {Niu}, \citenamefont {Ostby}, \citenamefont {Petukhov}, \citenamefont {Platt}, \citenamefont {Quintana}, \citenamefont {Rieffel}, \citenamefont {Roushan}, \citenamefont {Rubin}, \citenamefont {Sank}, \citenamefont {Satzinger}, \citenamefont {Smelyanskiy}, \citenamefont {Sung}, \citenamefont {Trevithick}, \citenamefont {Vainsencher}, \citenamefont {Villalonga}, \citenamefont {White}, \citenamefont {Yao}, \citenamefont {Yeh}, \citenamefont {Zalcman}, \citenamefont {Neven},\ and\ \citenamefont {Martinis}}]{Arute2019}%
  \BibitemOpen
  \bibfield  {author} {\bibinfo {author} {\bibfnamefont {F.}~\bibnamefont {Arute}}, \bibinfo {author} {\bibfnamefont {K.}~\bibnamefont {Arya}}, \bibinfo {author} {\bibfnamefont {R.}~\bibnamefont {Babbush}}, \bibinfo {author} {\bibfnamefont {D.}~\bibnamefont {Bacon}}, \bibinfo {author} {\bibfnamefont {J.~C.}\ \bibnamefont {Bardin}}, \bibinfo {author} {\bibfnamefont {R.}~\bibnamefont {Barends}}, \bibinfo {author} {\bibfnamefont {R.}~\bibnamefont {Biswas}}, \bibinfo {author} {\bibfnamefont {S.}~\bibnamefont {Boixo}}, \bibinfo {author} {\bibfnamefont {F.~G. S.~L.}\ \bibnamefont {Brandao}}, \bibinfo {author} {\bibfnamefont {D.~A.}\ \bibnamefont {Buell}}, \bibinfo {author} {\bibfnamefont {B.}~\bibnamefont {Burkett}}, \bibinfo {author} {\bibfnamefont {Y.}~\bibnamefont {Chen}}, \bibinfo {author} {\bibfnamefont {Z.}~\bibnamefont {Chen}}, \bibinfo {author} {\bibfnamefont {B.}~\bibnamefont {Chiaro}}, \bibinfo {author} {\bibfnamefont {R.}~\bibnamefont {Collins}}, \bibinfo {author} {\bibfnamefont {W.}~\bibnamefont
  {Courtney}}, \bibinfo {author} {\bibfnamefont {A.}~\bibnamefont {Dunsworth}}, \bibinfo {author} {\bibfnamefont {E.}~\bibnamefont {Farhi}}, \bibinfo {author} {\bibfnamefont {B.}~\bibnamefont {Foxen}}, \bibinfo {author} {\bibfnamefont {A.}~\bibnamefont {Fowler}}, \bibinfo {author} {\bibfnamefont {C.}~\bibnamefont {Gidney}}, \bibinfo {author} {\bibfnamefont {M.}~\bibnamefont {Giustina}}, \bibinfo {author} {\bibfnamefont {R.}~\bibnamefont {Graff}}, \bibinfo {author} {\bibfnamefont {K.}~\bibnamefont {Guerin}}, \bibinfo {author} {\bibfnamefont {S.}~\bibnamefont {Habegger}}, \bibinfo {author} {\bibfnamefont {M.~P.}\ \bibnamefont {Harrigan}}, \bibinfo {author} {\bibfnamefont {M.~J.}\ \bibnamefont {Hartmann}}, \bibinfo {author} {\bibfnamefont {A.}~\bibnamefont {Ho}}, \bibinfo {author} {\bibfnamefont {M.}~\bibnamefont {Hoffmann}}, \bibinfo {author} {\bibfnamefont {T.}~\bibnamefont {Huang}}, \bibinfo {author} {\bibfnamefont {T.~S.}\ \bibnamefont {Humble}}, \bibinfo {author} {\bibfnamefont {S.~V.}\ \bibnamefont
  {Isakov}}, \bibinfo {author} {\bibfnamefont {E.}~\bibnamefont {Jeffrey}}, \bibinfo {author} {\bibfnamefont {Z.}~\bibnamefont {Jiang}}, \bibinfo {author} {\bibfnamefont {D.}~\bibnamefont {Kafri}}, \bibinfo {author} {\bibfnamefont {K.}~\bibnamefont {Kechedzhi}}, \bibinfo {author} {\bibfnamefont {J.}~\bibnamefont {Kelly}}, \bibinfo {author} {\bibfnamefont {P.~V.}\ \bibnamefont {Klimov}}, \bibinfo {author} {\bibfnamefont {S.}~\bibnamefont {Knysh}}, \bibinfo {author} {\bibfnamefont {A.}~\bibnamefont {Korotkov}}, \bibinfo {author} {\bibfnamefont {F.}~\bibnamefont {Kostritsa}}, \bibinfo {author} {\bibfnamefont {D.}~\bibnamefont {Landhuis}}, \bibinfo {author} {\bibfnamefont {M.}~\bibnamefont {Lindmark}}, \bibinfo {author} {\bibfnamefont {E.}~\bibnamefont {Lucero}}, \bibinfo {author} {\bibfnamefont {D.}~\bibnamefont {Lyakh}}, \bibinfo {author} {\bibfnamefont {S.}~\bibnamefont {Mandr{\`{a}}}}, \bibinfo {author} {\bibfnamefont {J.~R.}\ \bibnamefont {McClean}}, \bibinfo {author} {\bibfnamefont {M.}~\bibnamefont
  {McEwen}}, \bibinfo {author} {\bibfnamefont {A.}~\bibnamefont {Megrant}}, \bibinfo {author} {\bibfnamefont {X.}~\bibnamefont {Mi}}, \bibinfo {author} {\bibfnamefont {K.}~\bibnamefont {Michielsen}}, \bibinfo {author} {\bibfnamefont {M.}~\bibnamefont {Mohseni}}, \bibinfo {author} {\bibfnamefont {J.}~\bibnamefont {Mutus}}, \bibinfo {author} {\bibfnamefont {O.}~\bibnamefont {Naaman}}, \bibinfo {author} {\bibfnamefont {M.}~\bibnamefont {Neeley}}, \bibinfo {author} {\bibfnamefont {C.}~\bibnamefont {Neill}}, \bibinfo {author} {\bibfnamefont {M.~Y.}\ \bibnamefont {Niu}}, \bibinfo {author} {\bibfnamefont {E.}~\bibnamefont {Ostby}}, \bibinfo {author} {\bibfnamefont {A.}~\bibnamefont {Petukhov}}, \bibinfo {author} {\bibfnamefont {J.~C.}\ \bibnamefont {Platt}}, \bibinfo {author} {\bibfnamefont {C.}~\bibnamefont {Quintana}}, \bibinfo {author} {\bibfnamefont {E.~G.}\ \bibnamefont {Rieffel}}, \bibinfo {author} {\bibfnamefont {P.}~\bibnamefont {Roushan}}, \bibinfo {author} {\bibfnamefont {N.~C.}\ \bibnamefont {Rubin}},
  \bibinfo {author} {\bibfnamefont {D.}~\bibnamefont {Sank}}, \bibinfo {author} {\bibfnamefont {K.~J.}\ \bibnamefont {Satzinger}}, \bibinfo {author} {\bibfnamefont {V.}~\bibnamefont {Smelyanskiy}}, \bibinfo {author} {\bibfnamefont {K.~J.}\ \bibnamefont {Sung}}, \bibinfo {author} {\bibfnamefont {M.~D.}\ \bibnamefont {Trevithick}}, \bibinfo {author} {\bibfnamefont {A.}~\bibnamefont {Vainsencher}}, \bibinfo {author} {\bibfnamefont {B.}~\bibnamefont {Villalonga}}, \bibinfo {author} {\bibfnamefont {T.}~\bibnamefont {White}}, \bibinfo {author} {\bibfnamefont {Z.~J.}\ \bibnamefont {Yao}}, \bibinfo {author} {\bibfnamefont {P.}~\bibnamefont {Yeh}}, \bibinfo {author} {\bibfnamefont {A.}~\bibnamefont {Zalcman}}, \bibinfo {author} {\bibfnamefont {H.}~\bibnamefont {Neven}},\ and\ \bibinfo {author} {\bibfnamefont {J.~M.}\ \bibnamefont {Martinis}},\ }\bibfield  {title} {\bibinfo {title} {{Quantum supremacy using a programmable superconducting processor}},\ }\href {https://doi.org/10.1038/s41586-019-1666-5} {\bibfield
  {journal} {\bibinfo  {journal} {Nature}\ }\textbf {\bibinfo {volume} {574}},\ \bibinfo {pages} {505} (\bibinfo {year} {2019})}\BibitemShut {NoStop}%
\bibitem [{\citenamefont {Pogorelov}\ \emph {et~al.}(2021)\citenamefont {Pogorelov}, \citenamefont {Feldker}, \citenamefont {Marciniak}, \citenamefont {Postler}, \citenamefont {Jacob}, \citenamefont {Krieglsteiner}, \citenamefont {Podlesnic}, \citenamefont {Meth}, \citenamefont {Negnevitsky}, \citenamefont {Stadler}, \citenamefont {H{\"{o}}fer}, \citenamefont {W{\"{a}}chter}, \citenamefont {Lakhmanskiy}, \citenamefont {Blatt}, \citenamefont {Schindler},\ and\ \citenamefont {Monz}}]{Pogorelov2021}%
  \BibitemOpen
  \bibfield  {author} {\bibinfo {author} {\bibfnamefont {I.}~\bibnamefont {Pogorelov}}, \bibinfo {author} {\bibfnamefont {T.}~\bibnamefont {Feldker}}, \bibinfo {author} {\bibfnamefont {C.~D.}\ \bibnamefont {Marciniak}}, \bibinfo {author} {\bibfnamefont {L.}~\bibnamefont {Postler}}, \bibinfo {author} {\bibfnamefont {G.}~\bibnamefont {Jacob}}, \bibinfo {author} {\bibfnamefont {O.}~\bibnamefont {Krieglsteiner}}, \bibinfo {author} {\bibfnamefont {V.}~\bibnamefont {Podlesnic}}, \bibinfo {author} {\bibfnamefont {M.}~\bibnamefont {Meth}}, \bibinfo {author} {\bibfnamefont {V.}~\bibnamefont {Negnevitsky}}, \bibinfo {author} {\bibfnamefont {M.}~\bibnamefont {Stadler}}, \bibinfo {author} {\bibfnamefont {B.}~\bibnamefont {H{\"{o}}fer}}, \bibinfo {author} {\bibfnamefont {C.}~\bibnamefont {W{\"{a}}chter}}, \bibinfo {author} {\bibfnamefont {K.}~\bibnamefont {Lakhmanskiy}}, \bibinfo {author} {\bibfnamefont {R.}~\bibnamefont {Blatt}}, \bibinfo {author} {\bibfnamefont {P.}~\bibnamefont {Schindler}},\ and\ \bibinfo {author}
  {\bibfnamefont {T.}~\bibnamefont {Monz}},\ }\bibfield  {title} {\bibinfo {title} {{Compact Ion-Trap Quantum Computing Demonstrator}},\ }\href {https://doi.org/10.1103/PRXQuantum.2.020343} {\bibfield  {journal} {\bibinfo  {journal} {PRX Quantum}\ }\textbf {\bibinfo {volume} {2}},\ \bibinfo {pages} {020343} (\bibinfo {year} {2021})}\BibitemShut {NoStop}%
\bibitem [{\citenamefont {Bluvstein}\ \emph {et~al.}(2023)\citenamefont {Bluvstein}, \citenamefont {Evered}, \citenamefont {Geim}, \citenamefont {Li}, \citenamefont {Zhou}, \citenamefont {Manovitz}, \citenamefont {Ebadi}, \citenamefont {Cain}, \citenamefont {Kalinowski}, \citenamefont {Hangleiter}, \citenamefont {Bonilla~Ataides}, \citenamefont {Maskara}, \citenamefont {Cong}, \citenamefont {Gao}, \citenamefont {Sales~Rodriguez}, \citenamefont {Karolyshyn}, \citenamefont {Semeghini}, \citenamefont {Gullans}, \citenamefont {Greiner}, \citenamefont {Vuletić},\ and\ \citenamefont {Lukin}}]{Bluvstein_2023}%
  \BibitemOpen
  \bibfield  {author} {\bibinfo {author} {\bibfnamefont {D.}~\bibnamefont {Bluvstein}}, \bibinfo {author} {\bibfnamefont {S.~J.}\ \bibnamefont {Evered}}, \bibinfo {author} {\bibfnamefont {A.~A.}\ \bibnamefont {Geim}}, \bibinfo {author} {\bibfnamefont {S.~H.}\ \bibnamefont {Li}}, \bibinfo {author} {\bibfnamefont {H.}~\bibnamefont {Zhou}}, \bibinfo {author} {\bibfnamefont {T.}~\bibnamefont {Manovitz}}, \bibinfo {author} {\bibfnamefont {S.}~\bibnamefont {Ebadi}}, \bibinfo {author} {\bibfnamefont {M.}~\bibnamefont {Cain}}, \bibinfo {author} {\bibfnamefont {M.}~\bibnamefont {Kalinowski}}, \bibinfo {author} {\bibfnamefont {D.}~\bibnamefont {Hangleiter}}, \bibinfo {author} {\bibfnamefont {J.~P.}\ \bibnamefont {Bonilla~Ataides}}, \bibinfo {author} {\bibfnamefont {N.}~\bibnamefont {Maskara}}, \bibinfo {author} {\bibfnamefont {I.}~\bibnamefont {Cong}}, \bibinfo {author} {\bibfnamefont {X.}~\bibnamefont {Gao}}, \bibinfo {author} {\bibfnamefont {P.}~\bibnamefont {Sales~Rodriguez}}, \bibinfo {author} {\bibfnamefont
  {T.}~\bibnamefont {Karolyshyn}}, \bibinfo {author} {\bibfnamefont {G.}~\bibnamefont {Semeghini}}, \bibinfo {author} {\bibfnamefont {M.~J.}\ \bibnamefont {Gullans}}, \bibinfo {author} {\bibfnamefont {M.}~\bibnamefont {Greiner}}, \bibinfo {author} {\bibfnamefont {V.}~\bibnamefont {Vuletić}},\ and\ \bibinfo {author} {\bibfnamefont {M.~D.}\ \bibnamefont {Lukin}},\ }\bibfield  {title} {\bibinfo {title} {Logical quantum processor based on reconfigurable atom arrays},\ }\href {https://doi.org/10.1038/s41586-023-06927-3} {\bibfield  {journal} {\bibinfo  {journal} {Nature}\ }\textbf {\bibinfo {volume} {626}},\ \bibinfo {pages} {58–65} (\bibinfo {year} {2023})}\BibitemShut {NoStop}%
\bibitem [{\citenamefont {Flamini}\ \emph {et~al.}(2019)\citenamefont {Flamini}, \citenamefont {Spagnolo},\ and\ \citenamefont {Sciarrino}}]{Flamini2018a}%
  \BibitemOpen
  \bibfield  {author} {\bibinfo {author} {\bibfnamefont {F.}~\bibnamefont {Flamini}}, \bibinfo {author} {\bibfnamefont {N.}~\bibnamefont {Spagnolo}},\ and\ \bibinfo {author} {\bibfnamefont {F.}~\bibnamefont {Sciarrino}},\ }\bibfield  {title} {\bibinfo {title} {Photonic quantum information processing: {A} review},\ }\href {https://doi.org/10.1088/1361-6633/aad5b2} {\bibfield  {journal} {\bibinfo  {journal} {Reports Prog. Phys.}\ }\textbf {\bibinfo {volume} {82}},\ \bibinfo {pages} {016001} (\bibinfo {year} {2019})}\BibitemShut {NoStop}%
\bibitem [{\citenamefont {Ringbauer}\ \emph {et~al.}(2021)\citenamefont {Ringbauer}, \citenamefont {Meth}, \citenamefont {Postler}, \citenamefont {Stricker}, \citenamefont {Blatt}, \citenamefont {Schindler},\ and\ \citenamefont {Monz}}]{ringbauer2021universal}%
  \BibitemOpen
  \bibfield  {author} {\bibinfo {author} {\bibfnamefont {M.}~\bibnamefont {Ringbauer}}, \bibinfo {author} {\bibfnamefont {M.}~\bibnamefont {Meth}}, \bibinfo {author} {\bibfnamefont {L.}~\bibnamefont {Postler}}, \bibinfo {author} {\bibfnamefont {R.}~\bibnamefont {Stricker}}, \bibinfo {author} {\bibfnamefont {R.}~\bibnamefont {Blatt}}, \bibinfo {author} {\bibfnamefont {P.}~\bibnamefont {Schindler}},\ and\ \bibinfo {author} {\bibfnamefont {T.}~\bibnamefont {Monz}},\ }\bibfield  {title} {\bibinfo {title} {A universal qudit quantum processor with trapped ions},\ }\href@noop {} {\  (\bibinfo {year} {2021})},\ \Eprint {https://arxiv.org/abs/2109.06903} {arXiv:2109.06903} \BibitemShut {NoStop}%
\bibitem [{\citenamefont {Morvan}\ \emph {et~al.}(2021)\citenamefont {Morvan}, \citenamefont {Ramasesh}, \citenamefont {Blok}, \citenamefont {Kreikebaum}, \citenamefont {O'Brien}, \citenamefont {Chen}, \citenamefont {Mitchell}, \citenamefont {Naik}, \citenamefont {Santiago},\ and\ \citenamefont {Siddiqi}}]{Morvan2020}%
  \BibitemOpen
  \bibfield  {author} {\bibinfo {author} {\bibfnamefont {A.}~\bibnamefont {Morvan}}, \bibinfo {author} {\bibfnamefont {V.~V.}\ \bibnamefont {Ramasesh}}, \bibinfo {author} {\bibfnamefont {M.~S.}\ \bibnamefont {Blok}}, \bibinfo {author} {\bibfnamefont {J.~M.}\ \bibnamefont {Kreikebaum}}, \bibinfo {author} {\bibfnamefont {K.}~\bibnamefont {O'Brien}}, \bibinfo {author} {\bibfnamefont {L.}~\bibnamefont {Chen}}, \bibinfo {author} {\bibfnamefont {B.~K.}\ \bibnamefont {Mitchell}}, \bibinfo {author} {\bibfnamefont {R.~K.}\ \bibnamefont {Naik}}, \bibinfo {author} {\bibfnamefont {D.~I.}\ \bibnamefont {Santiago}},\ and\ \bibinfo {author} {\bibfnamefont {I.}~\bibnamefont {Siddiqi}},\ }\bibfield  {title} {\bibinfo {title} {{Qutrit Randomized Benchmarking}},\ }\href {https://doi.org/10.1103/PhysRevLett.126.210504} {\bibfield  {journal} {\bibinfo  {journal} {Phys. Rev. Lett.}\ }\textbf {\bibinfo {volume} {126}},\ \bibinfo {pages} {210504} (\bibinfo {year} {2021})}\BibitemShut {NoStop}%
\bibitem [{\citenamefont {Chi}\ \emph {et~al.}(2022)\citenamefont {Chi}, \citenamefont {Huang}, \citenamefont {Zhang}, \citenamefont {Mao}, \citenamefont {Zhou}, \citenamefont {Chen}, \citenamefont {Zhai}, \citenamefont {Bao}, \citenamefont {Dai}, \citenamefont {Yuan}, \citenamefont {Zhang}, \citenamefont {Dai}, \citenamefont {Tang}, \citenamefont {Yang}, \citenamefont {Li}, \citenamefont {Ding}, \citenamefont {Oxenl{\o}we}, \citenamefont {Thompson}, \citenamefont {O'Brien}, \citenamefont {Li}, \citenamefont {Gong},\ and\ \citenamefont {Wang}}]{Chi2022}%
  \BibitemOpen
  \bibfield  {author} {\bibinfo {author} {\bibfnamefont {Y.}~\bibnamefont {Chi}}, \bibinfo {author} {\bibfnamefont {J.}~\bibnamefont {Huang}}, \bibinfo {author} {\bibfnamefont {Z.}~\bibnamefont {Zhang}}, \bibinfo {author} {\bibfnamefont {J.}~\bibnamefont {Mao}}, \bibinfo {author} {\bibfnamefont {Z.}~\bibnamefont {Zhou}}, \bibinfo {author} {\bibfnamefont {X.}~\bibnamefont {Chen}}, \bibinfo {author} {\bibfnamefont {C.}~\bibnamefont {Zhai}}, \bibinfo {author} {\bibfnamefont {J.}~\bibnamefont {Bao}}, \bibinfo {author} {\bibfnamefont {T.}~\bibnamefont {Dai}}, \bibinfo {author} {\bibfnamefont {H.}~\bibnamefont {Yuan}}, \bibinfo {author} {\bibfnamefont {M.}~\bibnamefont {Zhang}}, \bibinfo {author} {\bibfnamefont {D.}~\bibnamefont {Dai}}, \bibinfo {author} {\bibfnamefont {B.}~\bibnamefont {Tang}}, \bibinfo {author} {\bibfnamefont {Y.}~\bibnamefont {Yang}}, \bibinfo {author} {\bibfnamefont {Z.}~\bibnamefont {Li}}, \bibinfo {author} {\bibfnamefont {Y.}~\bibnamefont {Ding}}, \bibinfo {author} {\bibfnamefont {L.~K.}\
  \bibnamefont {Oxenl{\o}we}}, \bibinfo {author} {\bibfnamefont {M.~G.}\ \bibnamefont {Thompson}}, \bibinfo {author} {\bibfnamefont {J.~L.}\ \bibnamefont {O'Brien}}, \bibinfo {author} {\bibfnamefont {Y.}~\bibnamefont {Li}}, \bibinfo {author} {\bibfnamefont {Q.}~\bibnamefont {Gong}},\ and\ \bibinfo {author} {\bibfnamefont {J.}~\bibnamefont {Wang}},\ }\bibfield  {title} {\bibinfo {title} {{A programmable qudit-based quantum processor}},\ }\href {https://doi.org/10.1038/s41467-022-28767-x} {\bibfield  {journal} {\bibinfo  {journal} {Nat. Commun.}\ }\textbf {\bibinfo {volume} {13}},\ \bibinfo {pages} {1166} (\bibinfo {year} {2022})}\BibitemShut {NoStop}%
\bibitem [{\citenamefont {Lanyon}\ \emph {et~al.}(2008)\citenamefont {Lanyon}, \citenamefont {Barbieri}, \citenamefont {Almeida}, \citenamefont {Jennewein}, \citenamefont {Ralph}, \citenamefont {Resch}, \citenamefont {Pryde}, \citenamefont {O'Brien}, \citenamefont {Gilchrist},\ and\ \citenamefont {White}}]{Lanyon2008}%
  \BibitemOpen
  \bibfield  {author} {\bibinfo {author} {\bibfnamefont {B.~P.}\ \bibnamefont {Lanyon}}, \bibinfo {author} {\bibfnamefont {M.}~\bibnamefont {Barbieri}}, \bibinfo {author} {\bibfnamefont {M.~P.}\ \bibnamefont {Almeida}}, \bibinfo {author} {\bibfnamefont {T.}~\bibnamefont {Jennewein}}, \bibinfo {author} {\bibfnamefont {T.~C.}\ \bibnamefont {Ralph}}, \bibinfo {author} {\bibfnamefont {K.~J.}\ \bibnamefont {Resch}}, \bibinfo {author} {\bibfnamefont {G.~J.}\ \bibnamefont {Pryde}}, \bibinfo {author} {\bibfnamefont {J.~L.}\ \bibnamefont {O'Brien}}, \bibinfo {author} {\bibfnamefont {A.}~\bibnamefont {Gilchrist}},\ and\ \bibinfo {author} {\bibfnamefont {A.~G.}\ \bibnamefont {White}},\ }\bibfield  {title} {\bibinfo {title} {{Simplifying quantum logic using higher-dimensional Hilbert spaces}},\ }\href {https://doi.org/10.1038/nphys1150} {\bibfield  {journal} {\bibinfo  {journal} {Nat. Phys.}\ }\textbf {\bibinfo {volume} {5}},\ \bibinfo {pages} {134} (\bibinfo {year} {2008})}\BibitemShut {NoStop}%
\bibitem [{\citenamefont {Wang}\ \emph {et~al.}(2020)\citenamefont {Wang}, \citenamefont {Hu}, \citenamefont {Sanders},\ and\ \citenamefont {Kais}}]{Wang2020}%
  \BibitemOpen
  \bibfield  {author} {\bibinfo {author} {\bibfnamefont {Y.}~\bibnamefont {Wang}}, \bibinfo {author} {\bibfnamefont {Z.}~\bibnamefont {Hu}}, \bibinfo {author} {\bibfnamefont {B.~C.}\ \bibnamefont {Sanders}},\ and\ \bibinfo {author} {\bibfnamefont {S.}~\bibnamefont {Kais}},\ }\bibfield  {title} {\bibinfo {title} {Qudits and high-dimensional quantum computing},\ }\bibfield  {journal} {\bibinfo  {journal} {Frontiers in Physics}\ }\textbf {\bibinfo {volume} {8}},\ \href {https://doi.org/10.3389/fphy.2020.589504} {10.3389/fphy.2020.589504} (\bibinfo {year} {2020})\BibitemShut {NoStop}%
\bibitem [{\citenamefont {Ringbauer}\ \emph {et~al.}(2018)\citenamefont {Ringbauer}, \citenamefont {Bromley}, \citenamefont {Cianciaruso}, \citenamefont {Lami}, \citenamefont {Lau}, \citenamefont {Adesso}, \citenamefont {White}, \citenamefont {Fedrizzi},\ and\ \citenamefont {Piani}}]{Ringbauer2017}%
  \BibitemOpen
  \bibfield  {author} {\bibinfo {author} {\bibfnamefont {M.}~\bibnamefont {Ringbauer}}, \bibinfo {author} {\bibfnamefont {T.~R.}\ \bibnamefont {Bromley}}, \bibinfo {author} {\bibfnamefont {M.}~\bibnamefont {Cianciaruso}}, \bibinfo {author} {\bibfnamefont {L.}~\bibnamefont {Lami}}, \bibinfo {author} {\bibfnamefont {W.~Y.~S.}\ \bibnamefont {Lau}}, \bibinfo {author} {\bibfnamefont {G.}~\bibnamefont {Adesso}}, \bibinfo {author} {\bibfnamefont {A.~G.}\ \bibnamefont {White}}, \bibinfo {author} {\bibfnamefont {A.}~\bibnamefont {Fedrizzi}},\ and\ \bibinfo {author} {\bibfnamefont {M.}~\bibnamefont {Piani}},\ }\bibfield  {title} {\bibinfo {title} {{Certification and Quantification of Multilevel Quantum Coherence}},\ }\href {https://doi.org/10.1103/PhysRevX.8.041007} {\bibfield  {journal} {\bibinfo  {journal} {Phys. Rev. X}\ }\textbf {\bibinfo {volume} {8}},\ \bibinfo {pages} {041007} (\bibinfo {year} {2018})}\BibitemShut {NoStop}%
\bibitem [{\citenamefont {Kraft}\ \emph {et~al.}(2018)\citenamefont {Kraft}, \citenamefont {Ritz}, \citenamefont {Brunner}, \citenamefont {Huber},\ and\ \citenamefont {G{\"{u}}hne}}]{kraft2018}%
  \BibitemOpen
  \bibfield  {author} {\bibinfo {author} {\bibfnamefont {T.}~\bibnamefont {Kraft}}, \bibinfo {author} {\bibfnamefont {C.}~\bibnamefont {Ritz}}, \bibinfo {author} {\bibfnamefont {N.}~\bibnamefont {Brunner}}, \bibinfo {author} {\bibfnamefont {M.}~\bibnamefont {Huber}},\ and\ \bibinfo {author} {\bibfnamefont {O.}~\bibnamefont {G{\"{u}}hne}},\ }\bibfield  {title} {\bibinfo {title} {{Characterizing Genuine Multilevel Entanglement}},\ }\href {https://doi.org/10.1103/PhysRevLett.120.060502} {\bibfield  {journal} {\bibinfo  {journal} {Phys. Rev. Lett.}\ }\textbf {\bibinfo {volume} {120}},\ \bibinfo {pages} {060502} (\bibinfo {year} {2018})}\BibitemShut {NoStop}%
\bibitem [{\citenamefont {Gao}\ \emph {et~al.}(2023)\citenamefont {Gao}, \citenamefont {Appel}, \citenamefont {Friis}, \citenamefont {Ringbauer},\ and\ \citenamefont {Huber}}]{Gao2023}%
  \BibitemOpen
  \bibfield  {author} {\bibinfo {author} {\bibfnamefont {X.}~\bibnamefont {Gao}}, \bibinfo {author} {\bibfnamefont {P.}~\bibnamefont {Appel}}, \bibinfo {author} {\bibfnamefont {N.}~\bibnamefont {Friis}}, \bibinfo {author} {\bibfnamefont {M.}~\bibnamefont {Ringbauer}},\ and\ \bibinfo {author} {\bibfnamefont {M.}~\bibnamefont {Huber}},\ }\bibfield  {title} {\bibinfo {title} {On the role of entanglement in qudit-based circuit compression},\ }\href {https://doi.org/10.22331/q-2023-10-16-1141} {\bibfield  {journal} {\bibinfo  {journal} {{Quantum}}\ }\textbf {\bibinfo {volume} {7}},\ \bibinfo {pages} {1141} (\bibinfo {year} {2023})}\BibitemShut {NoStop}%
\bibitem [{\citenamefont {Gedik}\ \emph {et~al.}(2015)\citenamefont {Gedik}, \citenamefont {Silva}, \citenamefont {{\c{C}}akmak}, \citenamefont {Karpat}, \citenamefont {Vidoto}, \citenamefont {Soares-Pinto}, \citenamefont {DeAzevedo},\ and\ \citenamefont {Fanchini}}]{Gedik2015}%
  \BibitemOpen
  \bibfield  {author} {\bibinfo {author} {\bibfnamefont {Z.}~\bibnamefont {Gedik}}, \bibinfo {author} {\bibfnamefont {I.~A.}\ \bibnamefont {Silva}}, \bibinfo {author} {\bibfnamefont {B.}~\bibnamefont {{\c{C}}akmak}}, \bibinfo {author} {\bibfnamefont {G.}~\bibnamefont {Karpat}}, \bibinfo {author} {\bibfnamefont {E.~L.~G.}\ \bibnamefont {Vidoto}}, \bibinfo {author} {\bibfnamefont {D.~O.}\ \bibnamefont {Soares-Pinto}}, \bibinfo {author} {\bibfnamefont {E.~R.}\ \bibnamefont {DeAzevedo}},\ and\ \bibinfo {author} {\bibfnamefont {F.~F.}\ \bibnamefont {Fanchini}},\ }\bibfield  {title} {\bibinfo {title} {{Computational speed-up with a single qudit}},\ }\href {https://doi.org/10.1038/srep14671} {\bibfield  {journal} {\bibinfo  {journal} {Sci. Rep.}\ }\textbf {\bibinfo {volume} {5}},\ \bibinfo {pages} {14671} (\bibinfo {year} {2015})}\BibitemShut {NoStop}%
\bibitem [{\citenamefont {Zhan}\ \emph {et~al.}(2015)\citenamefont {Zhan}, \citenamefont {Li}, \citenamefont {Qin}, \citenamefont {hao Bian},\ and\ \citenamefont {Xue}}]{Zhan2015}%
  \BibitemOpen
  \bibfield  {author} {\bibinfo {author} {\bibfnamefont {X.}~\bibnamefont {Zhan}}, \bibinfo {author} {\bibfnamefont {J.}~\bibnamefont {Li}}, \bibinfo {author} {\bibfnamefont {H.}~\bibnamefont {Qin}}, \bibinfo {author} {\bibfnamefont {Z.}~\bibnamefont {hao Bian}},\ and\ \bibinfo {author} {\bibfnamefont {P.}~\bibnamefont {Xue}},\ }\bibfield  {title} {\bibinfo {title} {Linear optical demonstration of quantum speed-up with a single qudit},\ }\href {https://doi.org/10.1364/OE.23.018422} {\bibfield  {journal} {\bibinfo  {journal} {Optics Express}\ }\textbf {\bibinfo {volume} {23}},\ \bibinfo {pages} {18422} (\bibinfo {year} {2015})}\BibitemShut {NoStop}%
\bibitem [{\citenamefont {Meth}\ \emph {et~al.}(2024)\citenamefont {Meth}, \citenamefont {Haase}, \citenamefont {Zhang}, \citenamefont {Edmunds}, \citenamefont {Postler}, \citenamefont {Steiner}, \citenamefont {Jena}, \citenamefont {Dellantonio}, \citenamefont {Blatt}, \citenamefont {Zoller}, \citenamefont {Monz}, \citenamefont {Schindler}, \citenamefont {Muschik},\ and\ \citenamefont {Ringbauer}}]{meth2024simulating2dlatticegauge}%
  \BibitemOpen
  \bibfield  {author} {\bibinfo {author} {\bibfnamefont {M.}~\bibnamefont {Meth}}, \bibinfo {author} {\bibfnamefont {J.~F.}\ \bibnamefont {Haase}}, \bibinfo {author} {\bibfnamefont {J.}~\bibnamefont {Zhang}}, \bibinfo {author} {\bibfnamefont {C.}~\bibnamefont {Edmunds}}, \bibinfo {author} {\bibfnamefont {L.}~\bibnamefont {Postler}}, \bibinfo {author} {\bibfnamefont {A.}~\bibnamefont {Steiner}}, \bibinfo {author} {\bibfnamefont {A.~J.}\ \bibnamefont {Jena}}, \bibinfo {author} {\bibfnamefont {L.}~\bibnamefont {Dellantonio}}, \bibinfo {author} {\bibfnamefont {R.}~\bibnamefont {Blatt}}, \bibinfo {author} {\bibfnamefont {P.}~\bibnamefont {Zoller}}, \bibinfo {author} {\bibfnamefont {T.}~\bibnamefont {Monz}}, \bibinfo {author} {\bibfnamefont {P.}~\bibnamefont {Schindler}}, \bibinfo {author} {\bibfnamefont {C.}~\bibnamefont {Muschik}},\ and\ \bibinfo {author} {\bibfnamefont {M.}~\bibnamefont {Ringbauer}},\ }\href {https://arxiv.org/abs/2310.12110} {\bibinfo {title} {Simulating {2D} lattice gauge theories on a qudit
  quantum computer}} (\bibinfo {year} {2024}),\ \Eprint {https://arxiv.org/abs/2310.12110} {arXiv:2310.12110 [quant-ph]} \BibitemShut {NoStop}%
\bibitem [{\citenamefont {Edmunds}\ \emph {et~al.}(2024)\citenamefont {Edmunds}, \citenamefont {Rico}, \citenamefont {Arrazola}, \citenamefont {Brennen}, \citenamefont {Meth}, \citenamefont {Blatt},\ and\ \citenamefont {Ringbauer}}]{edmunds2024}%
  \BibitemOpen
  \bibfield  {author} {\bibinfo {author} {\bibfnamefont {C.~L.}\ \bibnamefont {Edmunds}}, \bibinfo {author} {\bibfnamefont {E.}~\bibnamefont {Rico}}, \bibinfo {author} {\bibfnamefont {I.}~\bibnamefont {Arrazola}}, \bibinfo {author} {\bibfnamefont {G.~K.}\ \bibnamefont {Brennen}}, \bibinfo {author} {\bibfnamefont {M.}~\bibnamefont {Meth}}, \bibinfo {author} {\bibfnamefont {R.}~\bibnamefont {Blatt}},\ and\ \bibinfo {author} {\bibfnamefont {M.}~\bibnamefont {Ringbauer}},\ }\href@noop {} {\bibinfo {title} {Constructing the spin-1 {{Haldane}} phase on a qudit quantum processor}} (\bibinfo {year} {2024})\BibitemShut {NoStop}%
\bibitem [{\citenamefont {Navickas}\ \emph {et~al.}(2024)\citenamefont {Navickas}, \citenamefont {MacDonell}, \citenamefont {Valahu}, \citenamefont {{Olaya-Agudelo}}, \citenamefont {Scuccimarra}, \citenamefont {Millican}, \citenamefont {Matsos}, \citenamefont {Nourse}, \citenamefont {Rao}, \citenamefont {Biercuk}, \citenamefont {Hempel}, \citenamefont {Kassal},\ and\ \citenamefont {Tan}}]{Navickas2024}%
  \BibitemOpen
  \bibfield  {author} {\bibinfo {author} {\bibfnamefont {T.}~\bibnamefont {Navickas}}, \bibinfo {author} {\bibfnamefont {R.~J.}\ \bibnamefont {MacDonell}}, \bibinfo {author} {\bibfnamefont {C.~H.}\ \bibnamefont {Valahu}}, \bibinfo {author} {\bibfnamefont {V.~C.}\ \bibnamefont {{Olaya-Agudelo}}}, \bibinfo {author} {\bibfnamefont {F.}~\bibnamefont {Scuccimarra}}, \bibinfo {author} {\bibfnamefont {M.~J.}\ \bibnamefont {Millican}}, \bibinfo {author} {\bibfnamefont {V.~G.}\ \bibnamefont {Matsos}}, \bibinfo {author} {\bibfnamefont {H.~L.}\ \bibnamefont {Nourse}}, \bibinfo {author} {\bibfnamefont {A.~D.}\ \bibnamefont {Rao}}, \bibinfo {author} {\bibfnamefont {M.~J.}\ \bibnamefont {Biercuk}}, \bibinfo {author} {\bibfnamefont {C.}~\bibnamefont {Hempel}}, \bibinfo {author} {\bibfnamefont {I.}~\bibnamefont {Kassal}},\ and\ \bibinfo {author} {\bibfnamefont {T.~R.}\ \bibnamefont {Tan}},\ }\href@noop {} {\bibinfo {title} {Experimental {{Quantum Simulation}} of {{Chemical Dynamics}}}} (\bibinfo {year} {2024})\BibitemShut
  {NoStop}%
\bibitem [{\citenamefont {Haldane}(1983)}]{Haldane1983}%
  \BibitemOpen
  \bibfield  {author} {\bibinfo {author} {\bibfnamefont {F.~D.~M.}\ \bibnamefont {Haldane}},\ }\bibfield  {title} {\bibinfo {title} {Nonlinear {{Field Theory}} of {{Large-Spin Heisenberg Antiferromagnets}}: {{Semiclassically Quantized Solitons}} of the {{One-Dimensional Easy-Axis N{\'e}el State}}},\ }\bibfield  {journal} {\bibinfo  {journal} {Physical Review Letters}\ }\textbf {\bibinfo {volume} {50}},\ \href {https://doi.org/10.1103/PhysRevLett.50.1153} {10.1103/PhysRevLett.50.1153} (\bibinfo {year} {1983})\BibitemShut {NoStop}%
\bibitem [{\citenamefont {Senko}\ \emph {et~al.}(2015)\citenamefont {Senko}, \citenamefont {Richerme}, \citenamefont {Smith}, \citenamefont {Lee}, \citenamefont {Cohen}, \citenamefont {Retzker},\ and\ \citenamefont {Monroe}}]{Senko2015}%
  \BibitemOpen
  \bibfield  {author} {\bibinfo {author} {\bibfnamefont {C.}~\bibnamefont {Senko}}, \bibinfo {author} {\bibfnamefont {P.}~\bibnamefont {Richerme}}, \bibinfo {author} {\bibfnamefont {J.}~\bibnamefont {Smith}}, \bibinfo {author} {\bibfnamefont {A.}~\bibnamefont {Lee}}, \bibinfo {author} {\bibfnamefont {I.}~\bibnamefont {Cohen}}, \bibinfo {author} {\bibfnamefont {A.}~\bibnamefont {Retzker}},\ and\ \bibinfo {author} {\bibfnamefont {C.}~\bibnamefont {Monroe}},\ }\bibfield  {title} {\bibinfo {title} {{Realization of a Quantum Integer-Spin Chain with Controllable Interactions}},\ }\href {https://doi.org/10.1103/PhysRevX.5.021026} {\bibfield  {journal} {\bibinfo  {journal} {Phys. Rev. X}\ }\textbf {\bibinfo {volume} {5}},\ \bibinfo {pages} {021026} (\bibinfo {year} {2015})}\BibitemShut {NoStop}%
\bibitem [{\citenamefont {Gonz\'alez-Cuadra}\ \emph {et~al.}(2022)\citenamefont {Gonz\'alez-Cuadra}, \citenamefont {Zache}, \citenamefont {Carrasco}, \citenamefont {Kraus},\ and\ \citenamefont {Zoller}}]{Cuadra2022}%
  \BibitemOpen
  \bibfield  {author} {\bibinfo {author} {\bibfnamefont {D.}~\bibnamefont {Gonz\'alez-Cuadra}}, \bibinfo {author} {\bibfnamefont {T.~V.}\ \bibnamefont {Zache}}, \bibinfo {author} {\bibfnamefont {J.}~\bibnamefont {Carrasco}}, \bibinfo {author} {\bibfnamefont {B.}~\bibnamefont {Kraus}},\ and\ \bibinfo {author} {\bibfnamefont {P.}~\bibnamefont {Zoller}},\ }\bibfield  {title} {\bibinfo {title} {{Hardware Efficient Quantum Simulation of Non-Abelian Gauge Theories with Qudits on Rydberg Platforms}},\ }\href {https://doi.org/10.1103/PhysRevLett.129.160501} {\bibfield  {journal} {\bibinfo  {journal} {Phys. Rev. Lett.}\ }\textbf {\bibinfo {volume} {129}},\ \bibinfo {pages} {160501} (\bibinfo {year} {2022})}\BibitemShut {NoStop}%
\bibitem [{\citenamefont {Calaj{\`o}}\ \emph {et~al.}(2024)\citenamefont {Calaj{\`o}}, \citenamefont {Magnifico}, \citenamefont {Edmunds}, \citenamefont {Ringbauer}, \citenamefont {Montangero},\ and\ \citenamefont {Silvi}}]{Calajo2024}%
  \BibitemOpen
  \bibfield  {author} {\bibinfo {author} {\bibfnamefont {G.}~\bibnamefont {Calaj{\`o}}}, \bibinfo {author} {\bibfnamefont {G.}~\bibnamefont {Magnifico}}, \bibinfo {author} {\bibfnamefont {C.}~\bibnamefont {Edmunds}}, \bibinfo {author} {\bibfnamefont {M.}~\bibnamefont {Ringbauer}}, \bibinfo {author} {\bibfnamefont {S.}~\bibnamefont {Montangero}},\ and\ \bibinfo {author} {\bibfnamefont {P.}~\bibnamefont {Silvi}},\ }\href@noop {} {\bibinfo {title} {{Digital Quantum Simulation of a (1+1){{D SU}}(2) Lattice Gauge Theory with Ion Qudits}}} (\bibinfo {year} {2024})\BibitemShut {NoStop}%
\bibitem [{\citenamefont {Popov}\ \emph {et~al.}(2024)\citenamefont {Popov}, \citenamefont {Kasper}, \citenamefont {Lewenstein}, \citenamefont {Zohar}, \citenamefont {Stornati},\ and\ \citenamefont {Hauke}}]{Popov2024}%
  \BibitemOpen
  \bibfield  {author} {\bibinfo {author} {\bibfnamefont {P.~P.}\ \bibnamefont {Popov}}, \bibinfo {author} {\bibfnamefont {V.}~\bibnamefont {Kasper}}, \bibinfo {author} {\bibfnamefont {M.}~\bibnamefont {Lewenstein}}, \bibinfo {author} {\bibfnamefont {E.}~\bibnamefont {Zohar}}, \bibinfo {author} {\bibfnamefont {P.}~\bibnamefont {Stornati}},\ and\ \bibinfo {author} {\bibfnamefont {P.}~\bibnamefont {Hauke}},\ }\bibfield  {title} {\bibinfo {title} {Non-perturbative signatures of fractons in the twisted multi-flavor {{Schwinger Model}}},\ }\href@noop {} {\bibfield  {journal} {\bibinfo  {journal} {Preprint at arXiv:2405.00745}\ } (\bibinfo {year} {2024})}\BibitemShut {NoStop}%
\bibitem [{\citenamefont {Wille}\ \emph {et~al.}(2024)\citenamefont {Wille}, \citenamefont {Berent}, \citenamefont {Forster}, \citenamefont {Kunasaikaran}, \citenamefont {Mato}, \citenamefont {Peham}, \citenamefont {Quetschlich}, \citenamefont {Rovara}, \citenamefont {Sander}, \citenamefont {Schmid}, \citenamefont {Schönberger}, \citenamefont {Stade},\ and\ \citenamefont {Burgholzer}}]{wille2024mqthandbooksummarydesign}%
  \BibitemOpen
  \bibfield  {author} {\bibinfo {author} {\bibfnamefont {R.}~\bibnamefont {Wille}}, \bibinfo {author} {\bibfnamefont {L.}~\bibnamefont {Berent}}, \bibinfo {author} {\bibfnamefont {T.}~\bibnamefont {Forster}}, \bibinfo {author} {\bibfnamefont {J.}~\bibnamefont {Kunasaikaran}}, \bibinfo {author} {\bibfnamefont {K.}~\bibnamefont {Mato}}, \bibinfo {author} {\bibfnamefont {T.}~\bibnamefont {Peham}}, \bibinfo {author} {\bibfnamefont {N.}~\bibnamefont {Quetschlich}}, \bibinfo {author} {\bibfnamefont {D.}~\bibnamefont {Rovara}}, \bibinfo {author} {\bibfnamefont {A.}~\bibnamefont {Sander}}, \bibinfo {author} {\bibfnamefont {L.}~\bibnamefont {Schmid}}, \bibinfo {author} {\bibfnamefont {D.}~\bibnamefont {Schönberger}}, \bibinfo {author} {\bibfnamefont {Y.}~\bibnamefont {Stade}},\ and\ \bibinfo {author} {\bibfnamefont {L.}~\bibnamefont {Burgholzer}},\ }\href {https://arxiv.org/abs/2405.17543} {\bibinfo {title} {{The {MQT} Handbook: A Summary of Design Automation Tools and Software for Quantum Computing}}} (\bibinfo
  {year} {2024}),\ \Eprint {https://arxiv.org/abs/2405.17543} {arXiv:2405.17543 [quant-ph]} \BibitemShut {NoStop}%
\bibitem [{\citenamefont {Clark}(2006)}]{Clark_2006}%
  \BibitemOpen
  \bibfield  {author} {\bibinfo {author} {\bibfnamefont {S.}~\bibnamefont {Clark}},\ }\bibfield  {title} {\bibinfo {title} {Valence bond solid formalism ford-level one-way quantum computation},\ }\href {https://doi.org/10.1088/0305-4470/39/11/010} {\bibfield  {journal} {\bibinfo  {journal} {Journal of Physics A: Mathematical and General}\ }\textbf {\bibinfo {volume} {39}},\ \bibinfo {pages} {2701–2721} (\bibinfo {year} {2006})}\BibitemShut {NoStop}%
\bibitem [{\citenamefont {Huber}\ and\ \citenamefont {de~Vicente}(2013)}]{Huber2013}%
  \BibitemOpen
  \bibfield  {author} {\bibinfo {author} {\bibfnamefont {M.}~\bibnamefont {Huber}}\ and\ \bibinfo {author} {\bibfnamefont {J.~I.}\ \bibnamefont {de~Vicente}},\ }\bibfield  {title} {\bibinfo {title} {Structure of multidimensional entanglement in multipartite systems},\ }\href {https://doi.org/10.1103/PhysRevLett.110.030501} {\bibfield  {journal} {\bibinfo  {journal} {Phys. Rev. Lett.}\ }\textbf {\bibinfo {volume} {110}},\ \bibinfo {pages} {030501} (\bibinfo {year} {2013})}\BibitemShut {NoStop}%
\bibitem [{\citenamefont {Cross}\ \emph {et~al.}(2017)\citenamefont {Cross}, \citenamefont {Bishop}, \citenamefont {Smolin},\ and\ \citenamefont {Gambetta}}]{cross2017openquantumassemblylanguage}%
  \BibitemOpen
  \bibfield  {author} {\bibinfo {author} {\bibfnamefont {A.~W.}\ \bibnamefont {Cross}}, \bibinfo {author} {\bibfnamefont {L.~S.}\ \bibnamefont {Bishop}}, \bibinfo {author} {\bibfnamefont {J.~A.}\ \bibnamefont {Smolin}},\ and\ \bibinfo {author} {\bibfnamefont {J.~M.}\ \bibnamefont {Gambetta}},\ }\href {https://arxiv.org/abs/1707.03429} {\bibinfo {title} {Open quantum assembly language}} (\bibinfo {year} {2017}),\ \Eprint {https://arxiv.org/abs/1707.03429} {arXiv:1707.03429 [quant-ph]} \BibitemShut {NoStop}%
\bibitem [{\citenamefont {Mato}\ \emph {et~al.}(2023{\natexlab{a}})\citenamefont {Mato}, \citenamefont {Ringbauer}, \citenamefont {Hillmich},\ and\ \citenamefont {Wille}}]{Mato23_Ent_Comp}%
  \BibitemOpen
  \bibfield  {author} {\bibinfo {author} {\bibfnamefont {K.}~\bibnamefont {Mato}}, \bibinfo {author} {\bibfnamefont {M.}~\bibnamefont {Ringbauer}}, \bibinfo {author} {\bibfnamefont {S.}~\bibnamefont {Hillmich}},\ and\ \bibinfo {author} {\bibfnamefont {R.}~\bibnamefont {Wille}},\ }\bibfield  {title} {\bibinfo {title} {Compilation of entangling gates for high-dimensional quantum systems},\ }in\ \href {https://doi.org/10.1145/3566097.3567930} {\emph {\bibinfo {booktitle} {Asia and South Pacific Design Automation Conf.}}},\ \bibinfo {series and number} {ASPDAC '23}\ (\bibinfo {year} {2023})\ p.\ \bibinfo {pages} {202–208}\BibitemShut {NoStop}%
\bibitem [{\citenamefont {Sawyer}\ and\ \citenamefont {Brown}(2021)}]{Sawyer2021}%
  \BibitemOpen
  \bibfield  {author} {\bibinfo {author} {\bibfnamefont {B.~C.}\ \bibnamefont {Sawyer}}\ and\ \bibinfo {author} {\bibfnamefont {K.~R.}\ \bibnamefont {Brown}},\ }\bibfield  {title} {\bibinfo {title} {{Wavelength-insensitive, multispecies entangling gate for group-2 atomic ions}},\ }\href {https://doi.org/10.1103/PhysRevA.103.022427} {\ \textbf {\bibinfo {volume} {103}},\ \bibinfo {pages} {022427} (\bibinfo {year} {2021})}\BibitemShut {NoStop}%
\bibitem [{\citenamefont {Hrmo}\ \emph {et~al.}(2022)\citenamefont {Hrmo}, \citenamefont {Wilhelm}, \citenamefont {Gerster}, \citenamefont {van Mourik}, \citenamefont {Huber}, \citenamefont {Blatt}, \citenamefont {Schindler}, \citenamefont {Monz},\ and\ \citenamefont {Ringbauer}}]{nativequdit}%
  \BibitemOpen
  \bibfield  {author} {\bibinfo {author} {\bibfnamefont {P.}~\bibnamefont {Hrmo}}, \bibinfo {author} {\bibfnamefont {B.}~\bibnamefont {Wilhelm}}, \bibinfo {author} {\bibfnamefont {L.}~\bibnamefont {Gerster}}, \bibinfo {author} {\bibfnamefont {M.~W.}\ \bibnamefont {van Mourik}}, \bibinfo {author} {\bibfnamefont {M.}~\bibnamefont {Huber}}, \bibinfo {author} {\bibfnamefont {R.}~\bibnamefont {Blatt}}, \bibinfo {author} {\bibfnamefont {P.}~\bibnamefont {Schindler}}, \bibinfo {author} {\bibfnamefont {T.}~\bibnamefont {Monz}},\ and\ \bibinfo {author} {\bibfnamefont {M.}~\bibnamefont {Ringbauer}},\ }\bibfield  {title} {\bibinfo {title} {Native qudit entanglement in a trapped ion quantum processor}\ }\href {https://doi.org/10.48550/ARXIV.2206.04104} {10.48550/ARXIV.2206.04104} (\bibinfo {year} {2022}),\ \Eprint {https://arxiv.org/abs/2206.04104} {arXiv:2206.04104} \BibitemShut {NoStop}%
\bibitem [{\citenamefont {Mato}\ \emph {et~al.}(2023{\natexlab{b}})\citenamefont {Mato}, \citenamefont {Hillmich},\ and\ \citenamefont {Wille}}]{MatoDD}%
  \BibitemOpen
  \bibfield  {author} {\bibinfo {author} {\bibfnamefont {K.}~\bibnamefont {Mato}}, \bibinfo {author} {\bibfnamefont {S.}~\bibnamefont {Hillmich}},\ and\ \bibinfo {author} {\bibfnamefont {R.}~\bibnamefont {Wille}},\ }\bibfield  {title} {\bibinfo {title} {Mixed-dimensional quantum circuit simulation with decision diagrams},\ }in\ \href {https://doi.org/10.1109/QCE57702.2023.00112} {\emph {\bibinfo {booktitle} {2023 IEEE International Conference on Quantum Computing and Engineering (QCE)}}},\ Vol.~\bibinfo {volume} {01}\ (\bibinfo {year} {2023})\ pp.\ \bibinfo {pages} {978--989}\BibitemShut {NoStop}%
\bibitem [{\citenamefont {Zulehner}\ and\ \citenamefont {Wille}(2019)}]{ZulehnerW19}%
  \BibitemOpen
  \bibfield  {author} {\bibinfo {author} {\bibfnamefont {A.}~\bibnamefont {Zulehner}}\ and\ \bibinfo {author} {\bibfnamefont {R.}~\bibnamefont {Wille}},\ }\bibfield  {title} {\bibinfo {title} {Advanced simulation of quantum computations},\ }\href {https://doi.org/10.1109/TCAD.2018.2834427} {\ \textbf {\bibinfo {volume} {38}},\ \bibinfo {pages} {848} (\bibinfo {year} {2019})}\BibitemShut {NoStop}%
\bibitem [{\citenamefont {Zulehner}\ \emph {et~al.}(2019)\citenamefont {Zulehner}, \citenamefont {Hillmich},\ and\ \citenamefont {Wille}}]{complex_ZulehnerHW19}%
  \BibitemOpen
  \bibfield  {author} {\bibinfo {author} {\bibfnamefont {A.}~\bibnamefont {Zulehner}}, \bibinfo {author} {\bibfnamefont {S.}~\bibnamefont {Hillmich}},\ and\ \bibinfo {author} {\bibfnamefont {R.}~\bibnamefont {Wille}},\ }\bibfield  {title} {\bibinfo {title} {How to efficiently handle complex values? {I}mplementing decision diagrams for quantum computing}\ }(\bibinfo  {publisher} {{ACM}},\ \bibinfo {year} {2019})\ pp.\ \bibinfo {pages} {1--7}\BibitemShut {NoStop}%
\bibitem [{\citenamefont {Niemann}\ \emph {et~al.}(2016)\citenamefont {Niemann}, \citenamefont {Wille}, \citenamefont {Miller}, \citenamefont {Thornton},\ and\ \citenamefont {Drechsler}}]{NiemannWMTD16}%
  \BibitemOpen
  \bibfield  {author} {\bibinfo {author} {\bibfnamefont {P.}~\bibnamefont {Niemann}}, \bibinfo {author} {\bibfnamefont {R.}~\bibnamefont {Wille}}, \bibinfo {author} {\bibfnamefont {D.~M.}\ \bibnamefont {Miller}}, \bibinfo {author} {\bibfnamefont {M.~A.}\ \bibnamefont {Thornton}},\ and\ \bibinfo {author} {\bibfnamefont {R.}~\bibnamefont {Drechsler}},\ }\bibfield  {title} {\bibinfo {title} {{QMDDs}: Efficient quantum function representation and manipulation},\ }\href@noop {} {\bibfield  {journal} {\bibinfo  {journal} {{IEEE} Trans. Comput. Aided Des. Integr. Circuits Syst.}\ }\textbf {\bibinfo {volume} {35}},\ \bibinfo {pages} {86} (\bibinfo {year} {2016})}\BibitemShut {NoStop}%
\bibitem [{\citenamefont {Roberts}\ \emph {et~al.}(2019)\citenamefont {Roberts}, \citenamefont {Milsted}, \citenamefont {Ganahl}, \citenamefont {Zalcman}, \citenamefont {Fontaine}, \citenamefont {Zou}, \citenamefont {Hidary}, \citenamefont {Vidal},\ and\ \citenamefont {Leichenauer}}]{roberts2019tensornetwork}%
  \BibitemOpen
  \bibfield  {author} {\bibinfo {author} {\bibfnamefont {C.}~\bibnamefont {Roberts}}, \bibinfo {author} {\bibfnamefont {A.}~\bibnamefont {Milsted}}, \bibinfo {author} {\bibfnamefont {M.}~\bibnamefont {Ganahl}}, \bibinfo {author} {\bibfnamefont {A.}~\bibnamefont {Zalcman}}, \bibinfo {author} {\bibfnamefont {B.}~\bibnamefont {Fontaine}}, \bibinfo {author} {\bibfnamefont {Y.}~\bibnamefont {Zou}}, \bibinfo {author} {\bibfnamefont {J.}~\bibnamefont {Hidary}}, \bibinfo {author} {\bibfnamefont {G.}~\bibnamefont {Vidal}},\ and\ \bibinfo {author} {\bibfnamefont {S.}~\bibnamefont {Leichenauer}},\ }\href@noop {} {\bibinfo {title} {Tensornetwork: A library for physics and machine learning}} (\bibinfo {year} {2019}),\ \Eprint {https://arxiv.org/abs/1905.01330} {arXiv:1905.01330 [physics.comp-ph]} \BibitemShut {NoStop}%
\bibitem [{\citenamefont {Orús}(2014)}]{Or_s_2014}%
  \BibitemOpen
  \bibfield  {author} {\bibinfo {author} {\bibfnamefont {R.}~\bibnamefont {Orús}},\ }\bibfield  {title} {\bibinfo {title} {A practical introduction to tensor networks: Matrix product states and projected entangled pair states},\ }\href {https://doi.org/10.1016/j.aop.2014.06.013} {\bibfield  {journal} {\bibinfo  {journal} {Annals of Physics}\ }\textbf {\bibinfo {volume} {349}},\ \bibinfo {pages} {117–158} (\bibinfo {year} {2014})}\BibitemShut {NoStop}%
\bibitem [{\citenamefont {Berquist}\ \emph {et~al.}(2022)\citenamefont {Berquist}, \citenamefont {Lykov}, \citenamefont {Liu},\ and\ \citenamefont {Alexeev}}]{Berquist}%
  \BibitemOpen
  \bibfield  {author} {\bibinfo {author} {\bibfnamefont {W.}~\bibnamefont {Berquist}}, \bibinfo {author} {\bibfnamefont {D.}~\bibnamefont {Lykov}}, \bibinfo {author} {\bibfnamefont {M.}~\bibnamefont {Liu}},\ and\ \bibinfo {author} {\bibfnamefont {Y.}~\bibnamefont {Alexeev}},\ }\bibfield  {title} {\bibinfo {title} {Stochastic approach for simulating quantum noise using tensor networks},\ }in\ \href {https://doi.org/10.1109/QCS56647.2022.00018} {\emph {\bibinfo {booktitle} {Third International Workshop on Quantum Computing Software (QCS)}}}\ (\bibinfo {year} {2022})\ pp.\ \bibinfo {pages} {107--113}\BibitemShut {NoStop}%
\bibitem [{\citenamefont {Grurl}\ \emph {et~al.}({\natexlab{a}})\citenamefont {Grurl}, \citenamefont {Kueng}, \citenamefont {Fu{\ss}},\ and\ \citenamefont {Wille}}]{GrurlKFW21}%
  \BibitemOpen
  \bibfield  {author} {\bibinfo {author} {\bibfnamefont {T.}~\bibnamefont {Grurl}}, \bibinfo {author} {\bibfnamefont {R.}~\bibnamefont {Kueng}}, \bibinfo {author} {\bibfnamefont {J.}~\bibnamefont {Fu{\ss}}},\ and\ \bibinfo {author} {\bibfnamefont {R.}~\bibnamefont {Wille}},\ }\bibfield  {title} {\bibinfo {title} {Stochastic quantum circuit simulation using decision diagrams},\ }in\ \href {https://doi.org/10.23919/DATE51398.2021.9474135} {\emph {\bibinfo {booktitle} {Design, Automation {\&} Test in Europe Conference}}},\ pp.\ \bibinfo {pages} {194--199}\BibitemShut {NoStop}%
\bibitem [{\citenamefont {Grurl}\ \emph {et~al.}({\natexlab{b}})\citenamefont {Grurl}, \citenamefont {Fuß},\ and\ \citenamefont {Wille}}]{noise_aware_grurl}%
  \BibitemOpen
  \bibfield  {author} {\bibinfo {author} {\bibfnamefont {T.}~\bibnamefont {Grurl}}, \bibinfo {author} {\bibfnamefont {J.}~\bibnamefont {Fuß}},\ and\ \bibinfo {author} {\bibfnamefont {R.}~\bibnamefont {Wille}},\ }\bibfield  {title} {\bibinfo {title} {Noise-aware quantum circuit simulation with decision diagrams},\ }\href {https://doi.org/10.1109/TCAD.2022.3182628} {\bibfield  {journal} {\bibinfo  {journal} {Transactions on Computer-Aided Design of Integrated Circuits and Systems}\ }\textbf {\bibinfo {volume} {42}},\ \bibinfo {pages} {860} ({\natexlab{b}})}\BibitemShut {NoStop}%
\bibitem [{\citenamefont {Harris}\ \emph {et~al.}(2020)\citenamefont {Harris}, \citenamefont {Millman}, \citenamefont {van~der Walt}, \citenamefont {Gommers}, \citenamefont {Virtanen}, \citenamefont {Cournapeau}, \citenamefont {Wieser}, \citenamefont {Taylor}, \citenamefont {Berg}, \citenamefont {Smith}, \citenamefont {Kern}, \citenamefont {Picus}, \citenamefont {Hoyer}, \citenamefont {van Kerkwijk}, \citenamefont {Brett}, \citenamefont {Haldane}, \citenamefont {del R{\'{i}}o}, \citenamefont {Wiebe}, \citenamefont {Peterson}, \citenamefont {G{\'{e}}rard-Marchant}, \citenamefont {Sheppard}, \citenamefont {Reddy}, \citenamefont {Weckesser}, \citenamefont {Abbasi}, \citenamefont {Gohlke},\ and\ \citenamefont {Oliphant}}]{harris2020array}%
  \BibitemOpen
  \bibfield  {author} {\bibinfo {author} {\bibfnamefont {C.~R.}\ \bibnamefont {Harris}}, \bibinfo {author} {\bibfnamefont {K.~J.}\ \bibnamefont {Millman}}, \bibinfo {author} {\bibfnamefont {S.~J.}\ \bibnamefont {van~der Walt}}, \bibinfo {author} {\bibfnamefont {R.}~\bibnamefont {Gommers}}, \bibinfo {author} {\bibfnamefont {P.}~\bibnamefont {Virtanen}}, \bibinfo {author} {\bibfnamefont {D.}~\bibnamefont {Cournapeau}}, \bibinfo {author} {\bibfnamefont {E.}~\bibnamefont {Wieser}}, \bibinfo {author} {\bibfnamefont {J.}~\bibnamefont {Taylor}}, \bibinfo {author} {\bibfnamefont {S.}~\bibnamefont {Berg}}, \bibinfo {author} {\bibfnamefont {N.~J.}\ \bibnamefont {Smith}}, \bibinfo {author} {\bibfnamefont {R.}~\bibnamefont {Kern}}, \bibinfo {author} {\bibfnamefont {M.}~\bibnamefont {Picus}}, \bibinfo {author} {\bibfnamefont {S.}~\bibnamefont {Hoyer}}, \bibinfo {author} {\bibfnamefont {M.~H.}\ \bibnamefont {van Kerkwijk}}, \bibinfo {author} {\bibfnamefont {M.}~\bibnamefont {Brett}}, \bibinfo {author} {\bibfnamefont
  {A.}~\bibnamefont {Haldane}}, \bibinfo {author} {\bibfnamefont {J.~F.}\ \bibnamefont {del R{\'{i}}o}}, \bibinfo {author} {\bibfnamefont {M.}~\bibnamefont {Wiebe}}, \bibinfo {author} {\bibfnamefont {P.}~\bibnamefont {Peterson}}, \bibinfo {author} {\bibfnamefont {P.}~\bibnamefont {G{\'{e}}rard-Marchant}}, \bibinfo {author} {\bibfnamefont {K.}~\bibnamefont {Sheppard}}, \bibinfo {author} {\bibfnamefont {T.}~\bibnamefont {Reddy}}, \bibinfo {author} {\bibfnamefont {W.}~\bibnamefont {Weckesser}}, \bibinfo {author} {\bibfnamefont {H.}~\bibnamefont {Abbasi}}, \bibinfo {author} {\bibfnamefont {C.}~\bibnamefont {Gohlke}},\ and\ \bibinfo {author} {\bibfnamefont {T.~E.}\ \bibnamefont {Oliphant}},\ }\bibfield  {title} {\bibinfo {title} {Array programming with {NumPy}},\ }\href {https://doi.org/10.1038/s41586-020-2649-2} {\bibfield  {journal} {\bibinfo  {journal} {Nature}\ }\textbf {\bibinfo {volume} {585}},\ \bibinfo {pages} {357} (\bibinfo {year} {2020})}\BibitemShut {NoStop}%
\bibitem [{\citenamefont {Mato}\ \emph {et~al.}(2024)\citenamefont {Mato}, \citenamefont {Hillmich},\ and\ \citenamefont {Wille}}]{Mato24_State_Prep}%
  \BibitemOpen
  \bibfield  {author} {\bibinfo {author} {\bibfnamefont {K.}~\bibnamefont {Mato}}, \bibinfo {author} {\bibfnamefont {S.}~\bibnamefont {Hillmich}},\ and\ \bibinfo {author} {\bibfnamefont {R.}~\bibnamefont {Wille}},\ }\bibfield  {title} {\bibinfo {title} {Mixed-dimensional qudit state preparation using edge-weighted decision diagrams},\ }\bibfield  {booktitle} {\emph {\bibinfo {booktitle} {Design Automation Conf.}},\ }\href@noop {} {\  (\bibinfo {year} {2024})}\BibitemShut {NoStop}%
\bibitem [{\citenamefont {Mato}\ \emph {et~al.}(2022)\citenamefont {Mato}, \citenamefont {Ringbauer}, \citenamefont {Hillmich},\ and\ \citenamefont {Wille}}]{Mato22_Single}%
  \BibitemOpen
  \bibfield  {author} {\bibinfo {author} {\bibfnamefont {K.}~\bibnamefont {Mato}}, \bibinfo {author} {\bibfnamefont {M.}~\bibnamefont {Ringbauer}}, \bibinfo {author} {\bibfnamefont {S.}~\bibnamefont {Hillmich}},\ and\ \bibinfo {author} {\bibfnamefont {R.}~\bibnamefont {Wille}},\ }\bibfield  {title} {\bibinfo {title} {Adaptive compilation of multi-level quantum operations},\ }in\ \href {https://doi.org/10.1109/QCE53715.2022.00070} {\emph {\bibinfo {booktitle} {International Conference on Quantum Computing and Engineering (QCE)}}}\ (\bibinfo {year} {2022})\ pp.\ \bibinfo {pages} {484--491}\BibitemShut {NoStop}%
\bibitem [{\citenamefont {Mato}\ \emph {et~al.}(2023{\natexlab{c}})\citenamefont {Mato}, \citenamefont {Hillmich},\ and\ \citenamefont {Wille}}]{Mato23_Compr}%
  \BibitemOpen
  \bibfield  {author} {\bibinfo {author} {\bibfnamefont {K.}~\bibnamefont {Mato}}, \bibinfo {author} {\bibfnamefont {S.}~\bibnamefont {Hillmich}},\ and\ \bibinfo {author} {\bibfnamefont {R.}~\bibnamefont {Wille}},\ }\bibfield  {title} {\bibinfo {title} {Compression of qubit circuits: Mapping to mixed-dimensional quantum systems},\ }in\ \href {https://doi.org/10.1109/QSW59989.2023.00027} {\emph {\bibinfo {booktitle} {International Conference on Quantum Software (QSW)}}}\ (\bibinfo {year} {2023})\ pp.\ \bibinfo {pages} {155--161}\BibitemShut {NoStop}%
\end{thebibliography}%


%

\end{document}